\documentclass[aps,pra,showpacs,amsfonts,notitlepage,amssymb,amsmath,nofootinbib,superscriptaddress]{revtex4-1}
\usepackage[utf8]{inputenc}
\usepackage[english]{babel}
\usepackage[titletoc,toc,title]{appendix}
\usepackage{amsmath}
\usepackage{amsfonts}
\usepackage{amssymb}
\usepackage[colorlinks=true,linkcolor=blue,citecolor=blue,urlcolor=blue]{hyperref}
\usepackage{graphicx}
\usepackage{enumerate}
\usepackage{csquotes}
\usepackage[caption=false]{subfig}
\usepackage{dsfont}
\usepackage{color}
\usepackage{bbold}
\usepackage{mathtools}
\usepackage{tensor}
\usepackage{gensymb}

\begin{document}

\title{Optical Kerr effect in vacuum}
\date{\today}

\author{Scott Robertson}
\affiliation{LAL, Univ. Paris-Sud, CNRS/IN2P3, Universit\'{e} Paris-Saclay, Orsay, France}

\begin{abstract}
From an effective field theory of electromagnetism in vacuum including all lowest-order nonlinear terms consistent with Lorentz invariance and locality of photon/photon interactions, 
we derive an effective medium description of strong background fields as regards their influence on a weak probe. 
We mainly consider as background a pump beam with well-defined wave vector and polarization. 
This leads us to define a nonlinear index of vacuum which, in the Euler-Heisenberg model derived from QED, 
has an optimal value of $1.555 \times 10^{-33} \, {\rm cm}^{2}/{\rm W}$ for a linearly polarized pump as seen by a counter-propagating, orthogonally polarized probe.  
We further generalize the model to include coupling to an axion field. 
In the limit where the axion mass is much smaller than the typical photon energy, 
this yields dispersive corrections, and the axionic signature is found to be greatly enhanced for a circularly polarized pump as compared to a linearly polarized one. 
The formalism here presented points to a simplification of the DeLLight experiment [X. Sarazin {\it et al.}, {\it Eur. Phys. J. D} {\bf 70}, 13 (2016)] aiming to measure the deflection of a probe by a tightly focused laser pulse.
\end{abstract}

\maketitle

\section{Introduction
\label{sec:introduction}}

In media, the dependence of optical properties on the electric or magnetic field 
has been known since the time of Faraday in the mid 19th-century~\cite{Faraday-1846-PhilMag}, though it has gained particular prominence in the last sixty years with the availability of high-intensity lasers and the subsequent development of nonlinear optics~\cite{Bloembergen-RMP,Bloembergen,Shen,Boyd,Agrawal}.
Such field-dependent behavior arises from the nontrivial response of bound charges and currents within the medium, leading to a range of nonlinear effects which have been extensively studied in the literature.
Since at optical frequencies %
the intensity associated with a single photon is 
very low, reaching the nonlinear regime requires 
an intense 
field comprising so many photons that it can be treated classically. 
From the point of view of a weak probe, a system comprising a strong background field and a 
medium 
can be treated 
as a single ``dressed'' medium, with the background field contributing to the total refractive index~\cite{Boyd, Agrawal}.
In typical dielectric media whose molecules possess 
inversion symmetry, the refractive index change is proportional to the square of the electric field, a phenomenon usually referred to as the {\it Kerr effect} after its discoverer John Kerr~\cite{Kerr-1875}.
When the index change 
is engendered by intense light (rather than static fields), %
this {\it optical Kerr effect} allows the assignment of a nonlinear index $n_{2}$ to the medium~\cite{Boyd,Agrawal} 
such that the total refractive index includes a term proportional to the intensity $I$ of the wave:
\begin{equation}
n(I) = n_{0} + n_{2} \, I \,,
\label{eq:Kerr_index_defn}
\end{equation}
where $n_{0}$ is the ``bare'' index of the medium %
in the absence of strong fields.
Though $n_{2}$ depends on the precise configuration (such as the relative polarization between the probe and the background), %
values 
typically range from $10^{-16}$ to $10^{-14} \, {\rm cm}^{2}/{\rm W}$ (see, e.g., Table 4.1.2 of~\cite{Boyd}).

In vacuum, by contrast, classical electrodynamics is a linear theory,
where $n_{2}$ is exactly zero. %
By analogy with the situation in media, we may ask whether this apparent linearity is only a low-field approximation, i.e., whether 
the field equations might become 
nonlinear when the fields are strong enough.
Indeed, the Standard Model already answers in the affirmative; in particular, quantum electrodynamics (QED) allows photon/photon scattering mediated by virtual electron/positron pairs, which play a role analogous to that of bound charges in media.
In the long-wavelength limit, 
this yields a nonlinear effective field theory for the electromagnetic field, the Lagrangian of which was %
first derived by Euler, Kockel and Heisenberg~\cite{Euler-Kockel-1935,Euler-1936,Heisenberg-Euler-1936}.
In principle, however, this is but one way in which nonlinearities could be generated: there may well be as-yet-unidentified particles (such as axions~\cite{Peccei-Quinn-1977-PRL,Peccei-Quinn-1977-PRD,Wilczek-1978,Weinberg-1978}) coupling to photons and thereby contributing 
to the effective nonlinear response; alternatively, there may be higher-order corrections to the classical electromagnetic sector of the Lagrangian (such as proposed by Born and Infeld~\cite{Born-Infeld-1934}).

It is thus of interest to experimentally probe nonlinear electrodynamics (NLED) in vacuum, in order to test our current predictions and potentially rule out alternative models~\cite{Battesti-Rizzo-2013,Fouche-Battesti-Rizzo-2016}.
While there have been several experiments in the high-energy photon regime which also tend to involve charged particles of some kind (a recent example is provided by the heavy ion collisions observed at the LHC~\cite{Atlas-2017}; see Ref.~\cite{Battesti-Rizzo-2013} for many others), the direct elastic scattering of real photons has not yet been observed, and the low-energy photon 
regime -- where, as in media, the strong background field is classical 
and nonlinearities can be treated 
as field-dependent contributions to the total refractive index -- remains relatively unexplored.
The most sensitive tests of low-energy NLED to date are those of the BMV~\cite{Battesti-et-al-2008,Cadene-et-al-2014} and PVLAS~\cite{Zavattini-et-al-2012,DellaValle-et-al-2016} experiments, which aim to detect the birefringence induced by a strong magnetic field perpendicular to the direction of the probe wave.
These have not yet reached 
the sensitivity required to test effects on the order of those predicted by QED.
An alternative, complementary approach is to directly exploit the intensity-dependence of the refractive index 
by using strong background fields to deflect 
the trajectory of a probe wave. 
This was attempted by Jones in 1960~\cite{Jones-1960,Jones-1961} using a static magnetic field, while the recently proposed DeLLight experiment~\cite{Sarazin-et-al-2016} aims to observe such a deflection using the much greater intensities within a tightly focused laser pulse.

Inspired by such proposals, we develop in this paper a theoretical framework for describing the lowest-order nonlinear interactions between electromagnetic waves in vacuum.
The novelty lies in the generality of the approach, which covers two aspects.
First, rather than restricting ourselves to the Euler-Heisenberg model describing the nonlinearities induced by QED, we consider instead a generalized model (in the spirit of Pleba\'{n}ski~\cite{Plebanski-Lectures} and Boillat~\cite{Boillat-1970}) which is consistent with Lorentz invariance and locality of effective photon/photon interactions.
As far as we are aware, this generalized Lagrangian was previously adopted only in~\cite{Bialynicka-Birula-1970} (for the particular case of static background fields), though the approach used in~\cite{Denisov-Denisova-2001-I} is reminiscent of it.
Moreover, being mindful of proposals for the detection of axions~\cite{Sikivie-1983,Maiani-Petronzio-Zavattini-1986,Gasperini-1987,vanBibber-et-al-1987,Raffelt-Stodolsky-1988}, we also allow a relaxation of the assumption of local photon/photon interactions through coupling to an axion-like field. 
This allows an explicit demonstration that, while the inclusion of the axion coupling would lead to a straightforward renormalization of the model parameters if the axion mass were large enough, the breakdown of locality of photon/photon interactions when this is not the case yields dispersive corrections to the behavior of a probe wave.
It also, in the case of an elliptically polarized background, induces some elliptical birefringence that is completely absent in the purely local theory.

The second aspect in which this paper generalizes the standard treatment of NLED concerns its explicitization of the effective medium engendered by an arbitrary configuration of strong background fields.
That is, assuming an explicit decomposition of the total fields into those of a strong background and a weak probe, the field equations can be linearized in those of the latter, in which case the ``dressed'' vacuum comprising the vacuum and strong background fields together behaves as an optical medium in its own right.
To this effective medium is assigned a set of well-defined susceptibility tensors. 
This description is more intuitive than the manifestly covariant formalism commonly used in the literature (as in, e.g.,~\cite{Brezin-Itzykson-1971}).
The general expression for the susceptibility tensors encompasses any background field configuration, describing both static and propagating fields; in the latter case, elliptical polarization is included in a relatively straightforward manner.
The behavior of a probe wave is completely determined by the optical properties encoded in the effective susceptibilities.
In particular, they allow a straightforward determination of the nonlinear 
index $n_{2}$ of vacuum by analogy with Eq.~(\ref{eq:Kerr_index_defn}). 
While a few previous works have given expressions for $n_{2}$ of vacuum~\cite{Aleksandrov-et-al-1985,Rozanov-1993}, they have worked with the Euler-Heisenberg model~\footnote{In the Euler-Heisenberg context, Eq.~(2.21) of~\cite{Rozanov-1993} gives a general result for propagating background fields that includes an arbitrary angle between the propagation directions of pump and probe, and even elliptical polarization, though how arbitrary polarization states enter is rather implicit.} rather than the generalized one adopted here,
and explicit values given~\cite{Boyd,Ferrando-et-al-2007} have only been estimates that lie somewhat below the optimized value~\footnote{Equation~(13.8.9) of~\cite{Boyd} gives only an estimate for $n_{2}$ as it neglects the tensorial nature of the response and is derived from a nonlinear permittivity that does not apply to weak probe waves on a strong background.  An order of magnitude is extracted from numerical simulations in~\cite{Ferrando-et-al-2007}, but not in the optimized scenario with the probe and background orthogonally polarized and counter-propagating.}.

The paper is organized as follows.  
In Sec.~\ref{sec:preliminaries} we lay the theoretical foundations by specifying the field normalization and Lagrangian we shall use, the latter being subject to the restrictions of Lorentz invariance and local interactions mentioned above.  We also indicate how particularly important models of NLED fit into this generalized framework.  
In Sec.~\ref{sec:effective_medium}, we develop the effective medium description by explicitly separating the total fields into a strong background and a weak probe, then linearizing the wave equations in the fields of the latter.  We pay particular attention to plane probe waves and their eigenstates within the effective medium, i.e., their refractive indices and polarizations, and we use this formalism to derive some known results in the case of static background fields.  
We consider an intense pump wave as a background in Sec.~\ref{sec:optical_Kerr}, showing how elliptical polarization of the pump can be taken into account and deriving the nonlinear 
index of vacuum by analogy with its definition in standard optical media.  
In Sec.~\ref{sec:axions}, we generalize the Lagrangian to include coupling to an axion field of arbitrary mass, yielding an effective theory of NLED which is nonlocal and thus characterized by dispersion.  
The analysis is carried through as before, and the key differences are emphasized.
We summarize our findings and conclude in Sec.~\ref{sec:conclusion}.

\section{Preliminaries 
\label{sec:preliminaries}}

We begin by establishing some theoretical foundations. 
First, we introduce a convenient normalization for the electromagnetic fields which simplifies the writing of many equations. 
We then state and discuss the most general form of the Lagrangian for the fields given a set of reasonable constraints.
This Lagrangian has three free parameters, and we finish this section by identifying the subsets of 
these parameters corresponding to 
two particularly important models of NLED: those of Euler-Heisenberg (EH) and Born-Infeld (BI). %

\subsection{Field normalization and equations} 

We work in Minkowski (flat) space, so that a 3-vector description of the electromagnetic fields may be straightforwardly applied. 
To avoid overuse of the fundamental constants $\epsilon_{0}$ and $\mu_{0}$ (respectively, the permittivity and permeability of free space), it is convenient to use the following rescaled definitions for the electric and magnetic fields:
\begin{equation}
{\bf E} = \sqrt{\epsilon_{0}} \, {\bf E}_{\rm SI} \,,  \qquad
{\bf D} = \frac{{\bf D}_{\rm SI}}{\sqrt{\epsilon_{0}}} \,,  \qquad
{\bf B} = \frac{{\bf B}_{\rm SI}}{\sqrt{\mu_{0}}} \,,  \qquad
{\bf H} = \sqrt{\mu_{0}} \, {\bf H}_{\rm SI} \,,
\label{eq:rescaling}
\end{equation}
where the subscript ``SI'' indicates the corresponding fields expressed in SI units.  With these definitions, each of the fields ${\bf E}$, ${\bf D}$, ${\bf B}$ and ${\bf H}$ has exactly the same units (the square root of an energy density), and the Maxwell equations in the absence of free charges and currents take the following form:
\begin{subequations}\begin{alignat}{2}
\nabla \cdot {\bf B} &= 0 \,, \qquad \nabla \times {\bf E} + \partial_{c t} {\bf B} \, &= 0 \,, \label{eq:Maxwell_a} \\
\nabla \cdot {\bf D} &= 0 \,, \qquad \nabla \times {\bf H} - \partial_{c t} {\bf D} \, &= 0 \,, \label{eq:Maxwell_b}
\end{alignat}
\label{eq:Maxwell}\end{subequations}
where $c = 1/\sqrt{\epsilon_{0} \mu_{0}}$ is the speed of light in vacuum.
Equations~(\ref{eq:Maxwell_a}) are automatically satisfied when ${\bf E}$ and ${\bf B}$ are defined in the standard covariant formulation as components of the antisymmetric tensor $F_{\mu\nu}$, itself defined as the exterior derivative of the four-potential $A_{\mu}$~\cite{Landau-Lifshitz-Fields}; in effect they are consistency conditions that allow such a writing to take place.  By contrast, Eqs.~(\ref{eq:Maxwell_b}) are a convenient writing of the Euler-Lagrange equations found by extremizing the action with respect to variations of $A_{\mu}$, where ${\bf D}$ and ${\bf H}$ are defined as
\begin{equation}
{\bf D} \doteq \frac{\partial\mathcal{L}}{\partial {\bf E}} \,, \qquad {\bf H} \doteq -\frac{\partial\mathcal{L}}{\partial {\bf B}} \,.
\label{eq:constitutive}
\end{equation}
While wave equations~(\ref{eq:Maxwell}) can only be fully solved once the constitutive equations relating 
${\bf D}$ and ${\bf H}$ to 
${\bf E}$ and ${\bf B}$ 
are specified, 
Eqs.~(\ref{eq:constitutive}) indicate that these relations are fully determined 
once the Lagrangian $\mathcal{L}\left({\bf E}\,,\,{\bf B}\right)$ is.

\subsection{Parametrized Lagrangian for NLED}

There are only two scalar quantities invariant under proper orthochronous Lorentz transformations (i.e., those continuously connected to the identity, 
requiring no 
spatial reflection or time reversal) which can be constructed from the electromagnetic fields alone~\cite{Landau-Lifshitz-Fields}:
\begin{equation}
\mathcal{F} \, \doteq \, -\frac{1}{4} F_{\mu\nu} F^{\mu\nu} \, = \, \frac{1}{2} \left( E^{2} - B^{2} \right) \,, \qquad 
\mathcal{G} \, \doteq \, -\frac{1}{4} F_{\mu\nu} \widetilde{F}^{\mu\nu} \, = \, {\bf E} \cdot {\bf B} \,,
\label{eq:scalars}
\end{equation}
where $\widetilde{F}^{\mu\nu} = \frac{1}{2} \epsilon^{\mu\nu\alpha\beta} F_{\alpha\beta}$ is the Hodge dual of $F_{\mu\nu}$, $\epsilon^{\mu\nu\alpha\beta}$ being the completely antisymmetric Levi-Civita symbol with $\epsilon^{0 1 2 3} = 1$. 
Therefore, a Lorentz invariant Lagrangian containing the  electromagnetic fields alone must depend only on $\mathcal{F}$ and $\mathcal{G}$, and if we further assume that the Lagrangian is spatially and temporally {\it local} (i.e., interactions are purely of the ``contact'' type), then its value at a given point of spacetime is straightforwardly a function of $\mathcal{F}$ and $\mathcal{G}$ at the same point: $\mathcal{L}\left(x\right) = \mathcal{L}\left(\mathcal{F}(x), \mathcal{G}(x)\right)$.  Such a Lagrangian for NLED is said to be of the Pleba\'{n}ski class~\cite{Plebanski-Lectures}.  We recognize $\mathcal{F}$ itself as the standard Lagrangian for electrodynamics in vacuum; it yields the trivial constitutive relations ${\bf D} = {\bf E}$ and ${\bf H} = {\bf B}$ when plugged into Eqs.~(\ref{eq:constitutive}), whence we recover the usual Maxwell equations in vacuum when inserted into Eqs.~(\ref{eq:Maxwell}).  Moreover, if $\mathcal{G}$ is added to the Lagrangian with some constant coefficient, it is straightforward to show that it has no effect on the field equations~\footnote{This is related to the fact that, when written in terms of $A_{\mu}$, $\mathcal{G}$ turns out to be a total derivative~\cite{Landau-Lifshitz-Fields}. \label{fn:total_derivative}}, and we are thus free to exclude the occurrence of such a term.  We thereby conclude that, at lowest order, the Lagrangian is simply $\mathcal{F}$.

In the weakly nonlinear regime, the Lagrangian can be expanded in powers of $\mathcal{F}$ and $\mathcal{G}$~\cite{Battesti-Rizzo-2013,Fouche-Battesti-Rizzo-2016}.  The first nonlinearities will be due to terms quadratic in $\mathcal{F}$ and $\mathcal{G}$ (i.e., quartic in the fields), and we parametrize their contribution as follows: 
\begin{equation}
\mathcal{L} = \mathcal{F} + \delta_{1} \, \frac{1}{2} \mathcal{F}^{2} + \delta_{2} \, \frac{1}{2} \mathcal{G}^{2} + \delta_{3} \, \mathcal{F} \mathcal{G}\, + ...
\label{eq:Lagrangian}
\end{equation}
The ``post-Maxwellian'' parameters~\cite{Denisov-Denisova-2001-I} 
$\delta_{1}$, $\delta_{2}$ and $\delta_{3}$ have units of inverse energy density.  They are not completely arbitrary: we will later see that they must satisfy certain inequalities in order for 
causality to be respected.  The term proportional to $\delta_{3}$ is often neglected as it breaks invariance under spatial reflection ($\mathcal{P}$) or time reversal ($\mathcal{T}$) transformations, which preserve the sign of $\mathcal{F}$ while causing $\mathcal{G}$ to flip sign.  Although QED is invariant under $\mathcal{P}$ and $\mathcal{T}$, the full Standard Model is not, $\mathcal{P}$-invariance being broken by weak interactions~\cite{Lee-Yang-1956,Wu-et-al-1957}.  Therefore, if we wish to include possible deviations from QED in our description, there is no fundamental reason why $\delta_{3}$ should vanish, and we keep it here for the sake of completeness.

Substituting into Eqs.~(\ref{eq:constitutive}), we find the nonlinear constitutive relations
\begin{equation}
{\bf D} = \frac{\partial\mathcal{L}}{\partial\mathcal{F}} \, {\bf E} + \frac{\partial\mathcal{L}}{\partial\mathcal{G}} \, {\bf B} \,, \qquad {\bf H} = \frac{\partial\mathcal{L}}{\partial\mathcal{F}} \, {\bf B} - \frac{\partial\mathcal{L}}{\partial\mathcal{G}} \, {\bf E}\,,
\label{eq:constitutive_NL_1}
\end{equation}
where
\begin{equation}
\frac{\partial\mathcal{L}}{\partial\mathcal{F}} = 1 + \delta_{1}\,\mathcal{F} + \delta_{3}\,\mathcal{G} + ... \,, \qquad \frac{\partial\mathcal{L}}{\partial\mathcal{G}} = \delta_{2}\,\mathcal{G} + \delta_{3}\,\mathcal{F} + ...
\label{eq:constitutive_NL_2}
\end{equation}
The first nonlinear 
terms in ${\bf D}$ and ${\bf H}$ %
are thus of third order in the fields, and can be thought of as defining the third-order susceptibilities of vacuum by analogy with 
media.  There is a key difference, of course: nonlinearities in media tend to be dominated 
by a single third-order susceptibility, governing the 
contribution to ${\bf D}$ which is cubic in the electric field~\footnote{In standard terminology, it is only this term that is referred to as the ``Kerr effect''.}. 
In vacuum, however, the requirement of Lorentz invariance restricts the possible nonlinearities to those described by Eqs.~(\ref{eq:constitutive_NL_1})-(\ref{eq:constitutive_NL_2}), and thus does not allow this 
term to occur on its own.
An important consequence of this difference is the following: since plane waves satisfy $\mathcal{F} = 0 = \mathcal{G} = 0$, they behave in vacuum NLED just as in the linear theory of Maxwell, with the same dispersion relation, $\omega = c k$.  
In the language of nonlinear optics, we may say that vacuum does not exhibit {\it self-phase modulation} (SPM), but only {\it cross-phase modulation} (XPM): interactions are induced only between different plane waves~\cite{Ferrando-et-al-2007}.  In 
media, on the other hand, SPM and XPM both occur, with different (though closely related) 
nonlinear indices (as defined in Eq.~(\ref{eq:Kerr_index_defn}))~\cite{Agrawal}.

\subsection{Euler-Heisenberg and Born-Infeld models}

Particular models yield particular values of and/or relations between the coefficients $\delta_{i}$.  We here consider two specific cases. %

The EH 
effective Lagrangian~\cite{Euler-1936,Heisenberg-Euler-1936} is derived from QED by summing over all Feynman diagrams containing a single electron-positron loop.  In performing the summation, it is assumed that the electromagnetic fields themselves are constant over the loop, which leads to a local effective theory~\footnote{This 
amounts to assuming that the typical photon wavelengths are much larger than the Compton wavelength of the electron,
which gives the characteristic size of the electron-positron loop. 
The loop can then be considered as a point-like vertex. 
Since $\lambda_{e} \sim 10^{-12} \, {\rm m}$, this is a good approximation at optical wavelengths $\gtrsim 10^{-7} \, {\rm m}$.}.  Therefore, when expanded to quartic order in the fields~\cite{Euler-Kockel-1935}, the EH Lagrangian takes the form~(\ref{eq:Lagrangian}), with $\delta_{3} = 0$ due to the $\mathcal{P}$- and $\mathcal{T}$-invariance of QED, and with the particular values
\begin{equation}
\delta_{1}^{\rm (EH)} = \frac{16}{45} \alpha^{2} \frac{\lambdabar_{e}^{3}}{m_{e} c^{2}} 
\approx 13.3\times 10^{-12} \, \mu{\rm m}^{3}/{\rm J}\,, \qquad \delta_{2}^{\rm (EH)} = \frac{7}{4} \delta_{1}^{\rm (EH)} \approx 23.3\times 10^{-12} \, \mu{\rm m}^{3}/{\rm J}\,.
\label{eq:EH-coeffs}
\end{equation}
Here, $\alpha \approx 1/137$ is the fine structure constant, $m_{e}$ is the mass of the electron, and $\lambdabar_{e} = \hbar/m_{e}c$ is the reduced Compton wavelength of the electron.  

The BI model~\cite{Born-Infeld-1934} 
is derived from the postulate that there exists a fundamental upper limit on the field strength, thus regularizing the self-energy of charged point particles.  This model is also $\mathcal{P}$- and $\mathcal{T}$-invariant so that $\delta_{3}=0$, but it predicts $\delta_{1} = \delta_{2} \doteq \delta^{\rm (BI)}$, in strict disagreement with the EH result given above.
It thus contains one free parameter, usually written as the maximum absolute field strength $b$, where $\delta^{\rm (BI)} = 1/b^{2}$.
No precise value is predicted, though 
Born and Infeld considered that the absolute field strength should be approximately that produced by an electron at its own classical radius, and using this prescription one finds 
\begin{equation}
\delta^{\rm (BI)} \sim 4 \pi \, \alpha^{3} \frac{\lambdabar_{e}^{3}}{m_{e} c^{2}} \approx 3.43 \times 10^{-12} \, \mu{\rm m}^{3}/{\rm J} \,,
\label{eq:BI-coeff}
\end{equation}
about a factor of $4$ smaller than $\delta_{1}^{\rm (EH)}$ (or a factor of $7$ smaller than $\delta_{2}^{\rm (EH)}$). 


\section{Effective medium description
\label{sec:effective_medium}}

In this section, we develop the analogy between the ``dressed'' vacuum (including strong electromagnetic fields) 
and an 
optical medium.  The Lagrangian and wave equations are explicitly decomposed into a background term describing the strong fields alone, and the lowest-order correction due to 
the presence of the probe.  A general equation for the probe wave eigenstates is derived, and some known results for the case of static 
background fields are reproduced. 

\subsection{Decomposition into background and probe fields
\label{sub:linearization}}

Much like in gravity, where we consider test particles assumed light enough not to have any significant effect on the gravitational field and whose motion is thus entirely determined by the spacetime metric already present, we wish here to consider probe waves propagating in a vacuum whose optical properties have been altered by the presence of strong fields, the probe waves being too weak to contribute to this alteration themselves. 
To this end, we decompose the total field into a sum of two terms: a background field, much the stronger of the two, entirely responsible for the alteration of the optical properties of the vacuum; and a significantly weaker probe field 
whose propagation through the altered vacuum we wish to solve for. 
The ``dressed'' vacuum (i.e., the combination of vacuum plus background fields) can be considered as a medium in its own right.
The insensitivity of the properties of this effective medium to the presence of the probe implies that the wave equations for the probe fields will be linear, or equivalently that 
the part of the Lagrangian relevant to the probe 
will be quadratic in those same fields.

Explicitly, let us write the total fields as ${\bf E} = {\bf E}_{0} + {\bf e}$ and ${\bf B} = {\bf B}_{0} + {\bf b}$, where ${\bf E}_{0}$ and ${\bf B}_{0}$ represent the background 
fields while ${\bf e}$ and ${\bf b}$ are the probe fields.  The Lagrangian is written as a Taylor series 
in the latter: 
\begin{equation}
\mathcal{L} = \mathcal{L}_{0}
+ \left. \frac{\partial\mathcal{L}}{\partial E_{i}} \right|_{0} \, e_{i} + \left. \frac{\partial\mathcal{L}}{\partial B_{i}} \right|_{0} \, b_{i}
+ \frac{1}{2} \left. \frac{\partial^{2}\mathcal{L}}{\partial E_{i} \partial E_{j}} \right|_{0} \, e_{i} e_{j} + \frac{1}{2} \left. \frac{\partial^{2}\mathcal{L}}{\partial B_{i} \partial B_{j}} \right|_{0} \, b_{i} b_{j} + \left. \frac{\partial^{2}\mathcal{L}}{\partial E_{i} \partial B_{j}} \right|_{0} \, e_{i} b_{j} + ...
\label{eq:Lagrangian_expansion}
\end{equation}
The subscript `0' indicates that the quantity in question is to be evaluated for 
the background fields ${\bf E}_{0}$ and ${\bf B}_{0}$,  which are taken to be solutions of the full nonlinear wave equations, extremizing the action by definition.  Therefore, the terms linear in ${\bf e}$ and ${\bf b}$ in Eq.~(\ref{eq:Lagrangian_expansion}), representing the first-order variation of $\mathcal{L}_{0}$, give zero contribution to the action and can 
be removed.  The first non-trivial terms involving the probe are those quadratic in the ${\bf e}$ and ${\bf b}$ fields, and 
we define this quadratic part as the effective Lagrangian for the probe:
\begin{equation}
\mathcal{L}_{\rm probe} \doteq \frac{1}{2} \left( e^{2} - b^{2} \right) + \frac{1}{2} {\bf e}^{\rm T} \delta\mathcal{L}_{\rm ee} {\bf e} + \frac{1}{2} {\bf b}^{\rm T} \delta\mathcal{L}_{\rm bb} {\bf b} + {\bf e}^{\rm T} \delta\mathcal{L}_{\rm eb} {\bf b} \,,
\label{eq:probe_Lagrangian}
\end{equation}
where the superscript `T' indicates the transpose, and where we have introduced the matrices
\begin{alignat}{6}
\left[ \delta\mathcal{L}_{\rm ee}\right]_{i j} & \doteq & \frac{\partial^{2}\left(\mathcal{L}-\mathcal{F}\right)}{\partial E_{i} \partial E_{j}} & = & \frac{\partial \left(D_{i} - E_{i}\right)}{\partial E_{j}} & = & \frac{\partial \left(D_{j} - E_{j}\right)}{\partial {E}_{i}} \,, \nonumber \\
\left[ \delta\mathcal{L}_{\rm bb}\right]_{i j} & \doteq & \,\, \frac{\partial^{2}\left(\mathcal{L}-\mathcal{F}\right)}{\partial B_{i} \partial B_{j}} & = & \frac{\partial \left(B_{i}-H_{i}\right)}{\partial B_{j}} & = & \frac{\partial \left(B_{j}-H_{j}\right)}{\partial B_{i}} \,, \nonumber \\
\left[ \delta\mathcal{L}_{\rm eb}\right]_{i j} & \doteq & \frac{\partial^{2}\left(\mathcal{L}-\mathcal{F}\right)}{\partial E_{i} \partial B_{j}} & = & \frac{\partial \left(D_{i}-E_{i}\right)}{\partial B_{j}} & = & \,\, \frac{\partial \left(B_{j}-H_{j}\right)}{\partial E_{i}} \,.
\label{eq:L-matrices}
\end{alignat}
These matrices clearly vanish when the full Lagrangian takes 
the standard Maxwell form (i.e., when $\mathcal{L} = \mathcal{F}$), in which case $\mathcal{L}_{\rm probe}$ is simply the Maxwell Lagrangian for the probe fields.  
Differentiating Eqs.~(\ref{eq:constitutive_NL_1}) as prescribed by the definitions in Eqs.~(\ref{eq:L-matrices}), we may write explicit expressions for the matrices $\delta\mathcal{L}_{\rm ee}$, $\delta\mathcal{L}_{\rm bb}$ and $\delta\mathcal{L}_{\rm eb}$: 
\begin{eqnarray}
\left[ \delta\mathcal{L}_{\rm ee}\right]_{i j} & = & \left( \delta_{1} \mathcal{F}_{0} + \delta_{3} \mathcal{G}_{0} \right) \delta_{i j} + \delta_{1} E_{0,i} E_{0,j} + \delta_{2} B_{0,i} B_{0,j} + \delta_{3} \left( E_{0,i} B_{0,j} + B_{0,i} E_{0,j} \right) \,, \nonumber \\
\left[ \delta\mathcal{L}_{\rm bb}\right]_{i j} & = & -\left( \delta_{1} \mathcal{F}_{0} + \delta_{3} \mathcal{G}_{0} \right) \delta_{i j} + \delta_{1} B_{0,i} B_{0,j} + \delta_{2} E_{0,i} E_{0,j} - \delta_{3} \left( E_{0,i} B_{0,j} + B_{0,i} E_{0,j} \right) \,, \nonumber \\
\left[ \delta\mathcal{L}_{\rm eb}\right]_{i j} & = & \left( \delta_{2} \mathcal{G}_{0} + \delta_{3} \mathcal{F}_{0} \right) \delta_{i j} - \delta_{1} E_{0,i} B_{0,j} + \delta_{2} B_{0,i} E_{0,j} + \delta_{3} \left( E_{0,i} E_{0,j} - B_{0,i} B_{0,j} \right) \,,
\label{eq:L-components}
\end{eqnarray}
where $\delta_{i j}$ is the Kronecker delta. 
Equivalently, using vector and matrix notation, we have
\begin{eqnarray}
\delta\mathcal{L}_{\rm ee} & = & \left( \delta_{1} \mathcal{F}_{0} + \delta_{3} \mathcal{G}_{0} \right) \mathbb{1} + \delta_{1} {\bf E}_{0} {\bf E}_{0}^{\rm T} + \delta_{2} {\bf B}_{0} {\bf B}_{0}^{\rm T} + \delta_{3} \left( {\bf E}_{0} {\bf B}_{0}^{\rm T} + {\bf B}_{0} {\bf E}_{0}^{\rm T} \right)  \,, \nonumber \\
\delta\mathcal{L}_{\rm bb} & = & -\left( \delta_{1} \mathcal{F}_{0} + \delta_{3} \mathcal{G}_{0} \right) \mathbb{1} + \delta_{1} {\bf B}_{0} {\bf B}_{0}^{\rm T} + \delta_{2} {\bf E}_{0} {\bf E}_{0}^{\rm T} - \delta_{3} \left( {\bf E}_{0} {\bf B}_{0}^{\rm T} + {\bf B}_{0} {\bf E}_{0}^{\rm T} \right) \,, \nonumber \\
\delta\mathcal{L}_{\rm eb} & = & \left( \delta_{2} \mathcal{G}_{0} + \delta_{3} \mathcal{F}_{0} \right) \mathbb{1} - \delta_{1} {\bf E}_{0} {\bf B}_{0}^{\rm T} + \delta_{2} {\bf B}_{0} {\bf E}_{0}^{\rm T} + \delta_{3} \left( {\bf E}_{0} {\bf E}_{0}^{\rm T} - {\bf B}_{0} {\bf B}_{0}^{\rm T} \right) \,,
\label{eq:L-vectors}
\end{eqnarray}
where $\mathbb{1}$ is the $3 \times 3$ identity matrix, and where the vector bi-products of the form ${\bf u} {\bf v}^{\rm T}$ are ``outer products'' yielding matrices rather than scalars.
Importantly, the $\delta\mathcal{L}$-matrices can be non-zero even when $\mathcal{F}_{0}$ and $\mathcal{G}_{0}$ vanish. 
Therefore, a single plane wave, though in some sense a ``linear'' solution of the wave equations, will nonetheless generate an effective medium as it will affect a probe wave whose propagation direction is not equal to its own~\footnote{In this respect, the original DeLLight proposal~\cite{Sarazin-et-al-2016} is over-complicated as it suggests using two counter-propagating pump beams to engender a nontrivial refractive index profile as seen by a probe.  One of the key points of this paper (examined further in Sec.~\ref{sec:optical_Kerr}) is that a single pump beam is sufficient for this purpose. \label{fn:DeLLight}}.

The probe Lagrangian~(\ref{eq:probe_Lagrangian}) allows us to treat ${\bf e}$ and ${\bf b}$ as the ``full'' electromagnetic fields, the background fields no longer being treated dynamically but rather having been subsumed into the definition of the effective medium. 
That is, Eqs.~(\ref{eq:Maxwell}) may be applied to the probe fields alone, and the associated constitutive relations are found by inserting $\mathcal{L}_{\rm probe}$ into Eqs.~(\ref{eq:constitutive}): 
\begin{equation}
{\bf d} = \left(\mathbb{1} + \overline{\delta\mathcal{L}}_{\rm ee}\right) {\bf e} + \overline{\delta\mathcal{L}}_{\rm eb} {\bf b} \,, \qquad {\bf h} = \left(\mathbb{1} - \overline{\delta\mathcal{L}}_{\rm bb}\right) {\bf b} - \overline{\delta\mathcal{L}}_{\rm eb}^{\rm T} {\bf e} \,.
\label{eq:probe_constitutive_1}
\end{equation}
Here, we have introduced an overbar on the $\delta\mathcal{L}$-matrices to indicate a spacetime average over the wavelength and period of the probe; equivalently, the overbar selects their ``slowly-varying'' component 
with respect to the oscillations of the probe.  
This ensures that ${\bf d}$ and ${\bf h}$ inherit the same carrier wave as ${\bf e}$ and ${\bf b}$, differing only in the form of their slowly-varying envelope. %
In general (and particularly when they are provided by a propagating wave) the background fields are highly oscillatory, and the $\delta\mathcal{L}$-matrices defined in Eqs.~(\ref{eq:L-components})-(\ref{eq:L-vectors}) will inherit some of this oscillatory behavior.  However, since the $\delta\mathcal{L}$-matrices influence the probe via an accumulated phase, 
the highly oscillatory terms 
can (to a good approximation) usually be neglected~\footnote{This ``rotating wave approximation'' is a standard procedure in nonlinear optics; see, e.g., Ref.~\cite{Agrawal}.  The rapidly oscillating terms in $\delta\mathcal{L}$ become significant only when {\it phase matching} occurs, i.e., when a certain combination of the wave vectors and frequencies involved generates another carrier %
wave which is itself 
``on-shell'', with frequency and wave vector approximately satisfying the dispersion relation $\omega = c k$.  For the quartic nonlinear Lagrangian considered here, there will be a total of four such waves in any given combination, and these processes are typically referred to as {\it four-wave mixing}.  See~\cite{Rozanov-1993,Moulin-Bernard-1999} for four-wave mixing in vacuum NLED.}.  

Rearranging Eqs.~(\ref{eq:probe_constitutive_1}), and neglecting products of the $\delta\mathcal{L}$-matrices (to be consistent with our neglect of higher-order terms in the Lagrangian~(\ref{eq:Lagrangian})), we find
\begin{equation}
{\bf d} = \left(\mathbb{1} + \overline{\delta\mathcal{L}}_{\rm ee}\right) {\bf e} + \overline{\delta\mathcal{L}}_{\rm eb} {\bf h} \,, \qquad {\bf b} = \left(\mathbb{1}+\overline{\delta\mathcal{L}}_{\rm bb}\right) {\bf h} + \overline{\delta\mathcal{L}}_{\rm eb}^{\rm T} {\bf e} \,.
\label{eq:probe_constitutive_2}
\end{equation}
Equations~(\ref{eq:probe_constitutive_2}) are in the standard form with respect to which the susceptibilities of an optical medium are defined. 
We may thus identify the effective 
electric, magnetic and magnetoelectric susceptibilities of the ``dressed'' vacuum: 
\begin{equation}
\chi^{\rm e} = \overline{\delta\mathcal{L}}_{\rm ee} \,, \qquad \chi^{\rm m} = \overline{\delta\mathcal{L}}_{\rm bb} \,, \qquad \alpha = \overline{\delta\mathcal{L}}_{\rm eb}\,.
\label{eq:susceptibilities}
\end{equation}

\subsection{Plane probe waves in the effective medium}

For definiteness, and without loss of generality, we take the probe to be propagating in the $-z$-direction.  The convenience of this choice stems from the fact that, when using a right-handed coordinate system, projections onto the $xy$-plane (with the $z$-axis pointing out of the page) intuitively represent what is ``seen'' by the probe during its propagation.  
Using the fact that the averaged $\overline{\delta\mathcal{L}}$-matrices are (by definition) slowly-varying with respect to the wavelength and period of the probe, we may locally decompose its electric field into a slowly-varying envelope and a carrier wave:
\begin{equation}
{\bf e}\left(z,t\right) = {\rm Re}\left\{ {\bf e}^{(0)} e^{-i k z - i\omega t} \right\} = \frac{1}{2} {\bf e}^{(0)} e^{-i k z - i\omega t} + {\rm c.c.}\,,
\end{equation}
where $k > 0$ and $\omega > 0$.  Analogous expressions hold for ${\bf b}$, ${\bf d}$ and ${\bf h}$.
The 
vector ${\bf e}^{(0)}$ determines the amplitude and polarization of the electric field, and likewise for ${\bf d}^{(0)}$, etc., while the ratio of $\omega$ to $k$ gives the phase velocity of the wave: $\omega/k = c/n$, where $n$ is the refractive index.  In the weakly nonlinear regime we are considering, 
$n$ will remain very close to $1$, in which case it is more convenient to express this relation in the form
\begin{equation}
1+\delta n = \frac{c k}{\omega} \,.
\label{eq:dn_definition}
\end{equation}
We wish to determine the refractive index variation $\delta n$, which satisfies $0 \leq \delta n \ll 1$~\footnote{The positivity of $\delta n$ stems from the requirement of causality in Special Relativity, i.e., that signals cannot propagate faster than the speed of light in vacuum, $c$.
Generally speaking, this applies not to the phase velocity $\omega/k$ but to the group velocity ${\rm d}\omega/{\rm d}k$, for which we may define a ``group index'' $n_{g} = n + \omega \, {\rm d}n/{\rm d}\omega$ such that the group velocity is $c/n_{g}$.
Then the causality condition is simply $n_{g} \geq 1$. 
In the present case, due to the local nature of the electromagnetic self-interaction encoded in the Lagrangian~(\ref{eq:Lagrangian}), $n$ will be independent of $\omega$, so the phase and group velocities are identical and this condition reduces to $\delta n \geq 0$. 
In Sec.~\ref{sec:axions}, we shall examine a dispersive case where $\delta n$ can be negative, yet the positivity of $\delta n_{g}$ is still respected. \label{fn:SR}} 
and which will (to lowest order) be quadratic in the background fields ${\bf E}_{0}$ and ${\bf B}_{0}$. 

Even before accounting for the constitutive equations~(\ref{eq:probe_constitutive_1}) relating ${\bf e}$ and ${\bf b}$ to ${\bf d}$ and ${\bf h}$, the probe must satisfy the independent set of equations~(\ref{eq:Maxwell}).  For a plane wave, these become
\begin{subequations}\begin{alignat}{2}
{\bf k} \cdot {\bf b}^{(0)} &= 0 \,, \qquad c\, {\bf k} \times {\bf e}^{(0)} - \omega \, {\bf b}^{(0)} &= 0 \,, \\
{\bf k} \cdot {\bf d}^{(0)} &= 0 \,, \qquad c\, {\bf k} \times {\bf h}^{(0)} + \omega \, {\bf d}^{(0)} &= 0 \,,
\end{alignat}
\label{eq:pw_Maxwell}\end{subequations}
where in the present case we have ${\bf k} = -k \, \hat{z}$.  
In each line of Eqs.~(\ref{eq:pw_Maxwell}), the second equation implies the first, so that these give only two independent equations rather than four.
Considering therefore only the second equation of each line, and using the definition of $\delta n$ given in Eq.~(\ref{eq:dn_definition}), Eqs.~(\ref{eq:pw_Maxwell}) may be written as
\begin{subequations}\begin{eqnarray}
{\bf b}^{(0)} & = & -\left(1 + \delta n\right) \Omega_{z} \, {\bf e}^{(0)} \,, \label{eq:pw_requirements_a}\\ 
{\bf d}^{(0)} & = & \left(1 + \delta n\right) \Omega_{z} \, {\bf h}^{(0)} \,, \label{eq:pw_requirements_b}
\end{eqnarray}
\label{eq:pw_requirements}\end{subequations}
where we have defined the $3 \times 3$ matrix
\begin{equation}
\Omega_{z} = \left[\begin{array}{ccc} 0 & -1 & 0 \\ 1 & 0 & 0 \\ 0 & 0 & 0 \end{array} \right]
\label{eq:Omega_definition}
\end{equation}
such that, when acting on a 3-dimensional vector ${\bf v}$, we have $\Omega_{z} {\bf v} = \hat{z} \times {\bf v}$.
We also note that, since $\Omega_{z}^{\rm T} = -\Omega_{z}$, the ordering is faithfully represented 
when acting on a transposed vector: ${\bf v}^{\rm T} \Omega_{z} = -\left(\Omega_{z} {\bf v}\right)^{\rm T} = -\left(\hat{z} \times {\bf v}\right)^{\rm T} = \left({\bf v} \times \hat{z}\right)^{\rm T}$.

It now remains to impose the constitutive relations~(\ref{eq:probe_constitutive_1}) for the probe.  
Using Eq.~(\ref{eq:pw_requirements_a}) so that both ${\bf d}^{(0)}$ and ${\bf h}^{(0)}$ can be written directly in terms of ${\bf e}^{(0)}$ (with no reference to ${\bf b}^{(0)}$), 
and neglecting products of small quantities, %
we have:
\begin{subequations}\begin{eqnarray}
{\bf d}^{(0)} & = & \left[ \mathbb{1} + \overline{\delta\mathcal{L}}_{\rm ee} - \overline{\delta\mathcal{L}}_{\rm eb} \Omega_{z} \right] {\bf e}^{(0)} \,, \label{eq:probe_constitutive_pw_a} \\
{\bf h}^{(0)} & = & \left[ -\left(1+\delta n\right) \Omega_{z} + \overline{\delta\mathcal{L}}_{\rm bb} \Omega_{z} - \overline{\delta\mathcal{L}}_{\rm eb}^{\rm T} \right] {\bf e}^{(0)} \,. \label{eq:probe_constitutive_pw_b}
\end{eqnarray}\label{eq:probe_constitutive_pw}\end{subequations}
Equations~(\ref{eq:probe_constitutive_pw}) are related to each other using Eq.~(\ref{eq:pw_requirements_b}),
thus yielding 
a single homogeneous equation involving the vector ${\bf e}^{(0)}$:  
\begin{equation}
\left[ \mathbb{1} + \Omega_{z}^{2} + 2 \, \delta n \, \Omega_{z}^{2} + \overline{\delta\mathcal{L}}_{\rm ee} - \Omega_{z} \, \overline{\delta\mathcal{L}}_{\rm bb} \, \Omega_{z} - \overline{\delta\mathcal{L}}_{\rm eb} \, \Omega_{z} - \left( \overline{\delta\mathcal{L}}_{\rm eb} \, \Omega_{z} \right)^{\rm T} \right] {\bf e}^{(0)} = 0 \,.
\label{eq:eigenvalue_eqn_3x3}
\end{equation}
The value of $\delta n$ is found by requiring the determinant of the operator in square brackets in Eq.~(\ref{eq:eigenvalue_eqn_3x3}) to vanish.  Note that, as $\Omega_{z}^{2} = {\rm diag}\left\{-1,-1,0\right\}$, $\delta n$ will appear only at quadratic order in the determinant (as could have been expected, since the two solutions correspond to the two possible polarizations of the probe wave).  
Moreover, since $\mathbb{1}+\Omega_{z}^{2} = {\rm diag}\left\{0,0,1\right\}$ and all other terms are of first order in $\delta n$ and the $\delta\mathcal{L}$-matrices, 
then to lowest nontrivial order only the $xy$-projection of Eq.~(\ref{eq:eigenvalue_eqn_3x3}) need be considered~\footnote{To see this, write Eq.~(\ref{eq:eigenvalue_eqn_3x3}) in the form $\left(\hat{z}\hat{z}^{\rm T} + \delta M\right) {\bf e}^{(0)} = 0$, where $\delta M$ is of lowest order in $\delta n$ and the $\delta\mathcal{L}$-matrices.  At zeroth-order, the longitudinal component $e^{(0)}_{z}$ vanishes, while the transverse component ${\bf e}^{(0)}_{\perp}$ is determined by the $xy$-projection of $\delta M$.}. 
Using expressions~(\ref{eq:L-vectors}) for the $\delta\mathcal{L}$-matrices 
and 
the above-mentioned 
identification of 
$\Omega_{z}$ with the cross product operator 
$\hat{z} \times $, Eq.~(\ref{eq:eigenvalue_eqn_3x3}) reduces to the following $2 \times 2$ eigenvalue problem:
\begin{equation}
\frac{1}{2} \left[ \delta_{1} \, \overline{\mathcal{E} \mathcal{E}^{\rm T}} + \delta_{2} \, \overline{\mathcal{B} \mathcal{B}^{\rm T}} + \delta_{3} \left( \overline{\mathcal{E} \mathcal{B}^{\rm T}} + \overline{\mathcal{B} \mathcal{E}^{\rm T}} \right) \right] {\bf e}^{(0)}_{\perp} = \delta n \, {\bf e}^{(0)}_{\perp} \,,
\label{eq:eigenvalue_eqn_2x2}
\end{equation}
where the subscript `$\perp$' indicates the projection onto the $xy$-plane, and where we have defined (in very similar fashion to Eq.~(20) of~\cite{Bialynicka-Birula-1970})
\begin{alignat}{4}
\mathcal{E} \,\, & = & \,\, {\bf E}_{0, \perp} - \hat{z} \times {\bf B}_{0, \perp} \,\, & = & \,\, -\hat{z} \times \left( \hat{z} \times {\bf E}_{0} \right) - \hat{z} \times {\bf B}_{0} \,, \nonumber \\
\mathcal{B} \,\, & = & \,\, {\bf B}_{0, \perp} + \hat{z} \times {\bf E}_{0, \perp} \,\, & = & \,\, -\hat{z} \times \left( \hat{z} \times {\bf B}_{0} \right) + \hat{z} \times {\bf E}_{0} \,.
\label{eq:vectors_defn}
\end{alignat}
It is clear that $\mathcal{E}$ and $\mathcal{B}$ lie in the $xy$-plane, and from their definition it immediately follows that $\mathcal{B} = \hat{z} \times \mathcal{E}$ and $\mathcal{E} = -\hat{z} \times \mathcal{B}$.  Therefore, for a given propagation direction, 
the behavior of the probe wave is determined 
by a single orthogonal vector 
formed from ${\bf E}_{0}$ and ${\bf B}_{0}$.

Equations~(\ref{eq:eigenvalue_eqn_2x2}) and~(\ref{eq:vectors_defn}) are one of the main results of this paper, giving the eigenstates of the probe in the effective medium generated by an arbitrary configuration of strong background fields.  They form the basis of the analysis up to the end of Sec.~\ref{sec:optical_Kerr} (before nonlocal corrections induced by an axion field are considered in Sec.~\ref{sec:axions}).

\subsection{Constant background fields: the DC Kerr and Cotton-Mouton effects
\label{sub:constantfields}}

As an illustrative example, we consider the simplest case where the background fields are constant, or at least slowly-varying with respect to the wavelength and period of the probe.  This case has already been analysed in some detail in the literature~\cite{Bialynicka-Birula-1970,Adler-1971,Rikken-Rizzo-2000,Rikken-Rizzo-2003}, though with particular emphasis on the predictions of the EH model.  We give a quick run-through of the various results here, showing that they are indeed reproduced by the effective medium framework we have used, and paving the way for the analysis of a propagating wave as background in Sec.~\ref{sec:optical_Kerr}.

\subsubsection{Refractive indices as eigenvalues}

Since the background fields are slowly-varying, the overbars in Eq.~(\ref{eq:eigenvalue_eqn_2x2}) are redundant, the $2 \times 2$ eigenvalue equation reducing to
\begin{equation}
\frac{1}{2} \left[ \delta_{1} \, \mathcal{E} \mathcal{E}^{\rm T} + \delta_{2} \, \mathcal{B} \mathcal{B}^{\rm T} + \delta_{3} \left( \mathcal{E} \mathcal{B}^{\rm T} + \mathcal{B} \mathcal{E}^{\rm T} \right) \right] {\bf e}^{(0)}_{\perp} = \delta n \, {\bf e}^{(0)}_{\perp} \,,
\label{eq:eigenvalue_eqn_2x2_constantfields}
\end{equation}
where the vectors $\mathcal{E}$ and $\mathcal{B}$ are now to be considered as constant.
Since $\mathcal{E}$ and $\mathcal{B}$ have the same magnitude and are perpendicular to each other, we may use the orthonormal vectors 
$\hat{\mathcal{E}} = \mathcal{E}/\left|\mathcal{E}\right|$ and $\hat{\mathcal{B}} = \mathcal{B}/\left|\mathcal{B}\right|$ as a basis in the $xy$-plane. 
The ordered vectors $\left\{ \hat{\mathcal{E}} \,,\, \hat{\mathcal{B}} \,,\, \hat{z} \right\}$ form a right-handed orthonormal basis, 
$\hat{\mathcal{E}}$ and $\hat{\mathcal{B}}$ being 
analogous to the standard Cartesian basis vectors 
$\hat{x}$ and $\hat{y}$, respectively.  In the $\left\{\hat{\mathcal{E}} \,,\, \hat{\mathcal{B}} \right\}$ basis, Eq.~(\ref{eq:eigenvalue_eqn_2x2_constantfields}) can be written in matrix notation as
\begin{equation}
\Bigg[ \begin{array}{cc} \delta_{1} & \delta_{3} \\ \delta_{3} & \delta_{2} \end{array} \Bigg] \left[ \begin{array}{c} e^{(0)}_{\mathcal{E}} \\ e^{(0)}_{\mathcal{B}} \end{array} \right] = \frac{\delta n}{\frac{1}{2} \left|\mathcal{E}\right|^{2}} \, \left[ \begin{array}{c} e^{(0)}_{\mathcal{E}} \\ e^{(0)}_{\mathcal{B}} \end{array} \right] \,.
\label{eq:eigenproblem_constant}
\end{equation}
This is readily solved.  
There are two refractive indices (corresponding to two polarizations of the probe wave),
\begin{equation}
\delta n_{\pm} = \delta_{\pm} \cdot \frac{1}{2} \left| \mathcal{E} \right|^{2} \,,
\label{eq:delta-n_constant}
\end{equation}
where the coefficients $\delta_{\pm}$ are the eigenvalues of the matrix on the left-hand side of Eq.~(\ref{eq:eigenproblem_constant}):
\begin{equation}
\delta_{\pm} = \frac{1}{2}\left( \delta_{1} + \delta_{2} \pm \sqrt{\left(\delta_{1}-\delta_{2}\right)^{2} + \left(2 \delta_{3}\right)^{2}} \right) \,.
\label{eq:delta_constant}
\end{equation}
By definition, $\delta_{+} \geq \delta_{-}$ and $\delta n_{+} \geq \delta n_{-}$.  Equality only holds when 
$\delta_{1}-\delta_{2} = \delta_{3} = 0$ (as in the BI model); otherwise 
the presence of the background fields makes the vacuum birefringent~\cite{Rikken-Rizzo-2000}, a phenomenon referred to as the {\it DC Kerr effect} when the static external field is a pure electric field and the {\it Cotton-Mouton effect} when it is a pure magnetic field.  
The vacuum Cotton-Mouton effect is the basis for the BMV~\cite{Battesti-et-al-2008,Cadene-et-al-2014} and PVLAS~\cite{Zavattini-et-al-2012,DellaValle-et-al-2016} experiments.  For a magnetic field oriented perpendicular to the direction of the probe wave, 
we have $\left| \mathcal{E} \right|^{2} = B_{0}^{2} = B_{0, {\rm SI}}^{2}/\mu_{0}$, 
and the difference in the two refractive indices is 
\begin{equation}
\Delta n = \sqrt{\left(\delta_{1}-\delta_{2}\right)^2 + \left(2 \delta_{3}\right)^{2}} \, \frac{B_{0,{\rm SI}}^{2}}{2\mu_{0}} \,.
\end{equation}
For a magnetic field of 1 Tesla, this gives, for the EH and BI models,
\begin{equation}
\Delta n^{\rm (EH)} = 3.98 \times 10^{-24} \,, \qquad \Delta n^{\rm (BI)} = 0 \,.
\end{equation}
The EH value is in agreement with the predictions of Refs.~\cite{Battesti-et-al-2008,Cadene-et-al-2014,Zavattini-et-al-2012,DellaValle-et-al-2016}, while the vanishing of $\Delta n^{\rm (BI)}$ indicates the absence of birefringence in the BI model~\cite{Boillat-1970,Plebanski-Lectures,Bialynicka-Birula-1970}.

As noted in footnote~\ref{fn:SR}, respecting 
causality requires the avoidance of a negative value of $\delta n_{\pm}$, or equivalently of $\delta_{\pm}$.  It is straightforward to show that this implies the inequalities $\delta_{1} \geq 0$, $\delta_{2} \geq 0$ and $\delta_{1}\delta_{2} - \delta_{3}^{2} \geq 0$ (in agreement with Eqs.~(25) of~\cite{Bialynicka-Birula-1970}).  Interestingly, using the identification of the effective susceptibilities made in Eqs.~(\ref{eq:susceptibilities}), and using the simplifying assumption $\mathcal{F}_{0} = \mathcal{G}_{0} = 0$, we find 
\begin{equation}
\chi^{\rm e}_{ii} \chi^{\rm m}_{jj} - \left(\alpha_{i j}\right)^{2} = \left(\delta_{1}\delta_{2}-\delta_{3}^{2}\right) \left(E_{0,i} E_{0,j} + B_{0,i} B_{0,j}\right)^{2} \,.
\end{equation}
So in this case the inequality $\delta_{1}\delta_{2}-\delta_{3}^{2} \geq 0$, derived here from the requirement of causality, is equivalent to $\chi^{\rm e}_{ii} \chi^{\rm m}_{jj} - \left(\alpha_{i j}\right)^{2} \geq 0$, 
previously derived (for any optical medium) %
in~\cite{Brown-Hornreich-Shtrikman-1968} from the requirement of thermodynamic stability.

\subsubsection{Anisotropy of the effective medium}

While the factors $\delta_{\pm}$ are fixed by the post-Maxwellian parameters entering Eq.~(\ref{eq:Lagrangian}), the strength of the refractive index change is also proportional to $\frac{1}{2}\left|\mathcal{E}\right|^{2}$.  This is simply quadratic in the background fields, but because of the projection and combination required to form $\mathcal{E}$ and $\mathcal{B}$, the dependence on relative orientation (between ${\bf E}_{0}$ and ${\bf B}_{0}$, as well as between these fields and the probe wave vector ${\bf k} \propto -\hat{z}$) can be rather complicated.  After a bit of algebra, it can be shown that 
\begin{equation}
\frac{1}{2} \left|\mathcal{E}\right|^{2} = \frac{1}{2} \left( \left| \hat{z} \times {\bf E}_{0} \right|^{2} + \left| \hat{z} \times {\bf B}_{0} \right|^{2} \right) + \hat{z} \cdot \left( {\bf E}_{0} \times {\bf B}_{0} \right) \,.
\end{equation}
The first term here is rather simple, in that ${\bf E}_{0}$ and ${\bf B}_{0}$ contribute separately, and with the squared magnitude of their projections onto the $xy$-plane.  The second term is more subtle, as it depends on the relative orientation of ${\bf E}_{0}$ and ${\bf B}_{0}$.  Moreover, it is directionally dependent: whereas the first term depends only on the line along which the wave vector ${\bf k}$ lies (defined to be the $z$-axis) and does not vary under the transformation ${\bf k} \to -{\bf k}$ (i.e., $\hat{z} \to -\hat{z}$), the second term changes sign under this transformation.  In this sense the effective medium behaves as if it were moving with a velocity proportional to the ``Poynting vector'' ${\bf E}_{0} \times {\bf B}_{0}$.  This anisotropy (which is independent of the probe polarization as it stems only from the magnitude of the vector $\mathcal{E}$) was described in Ref.~\cite{Rikken-Rizzo-2003}.  We shall see that it is also present when the background is a plane wave, with co-propagating probe waves seeing no refractive index change while counter-propagating waves experience the strongest effect.

\subsubsection{Eigenpolarizations}

\begin{figure}
\includegraphics[width=0.45\columnwidth]{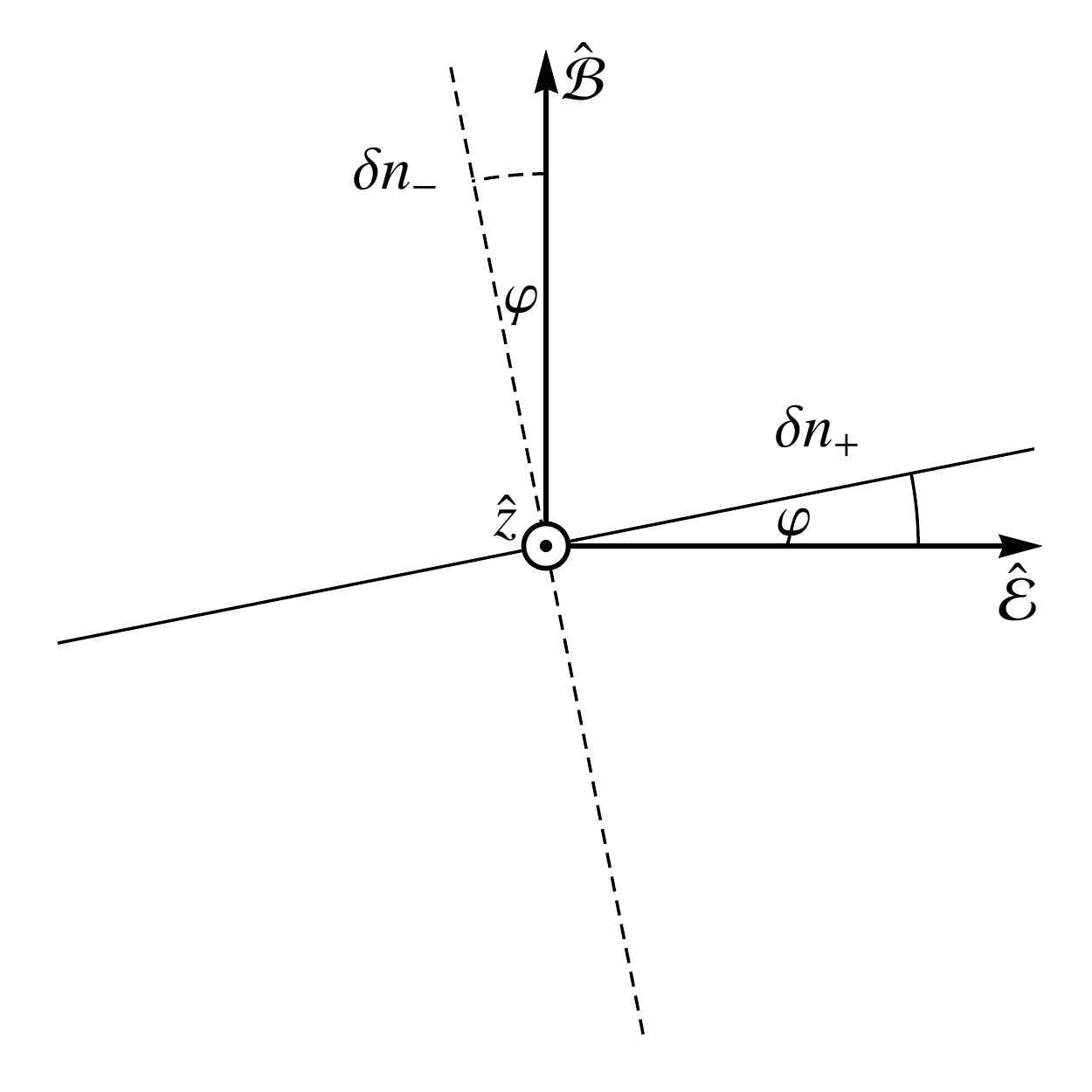} \, \includegraphics[width=0.45\columnwidth]{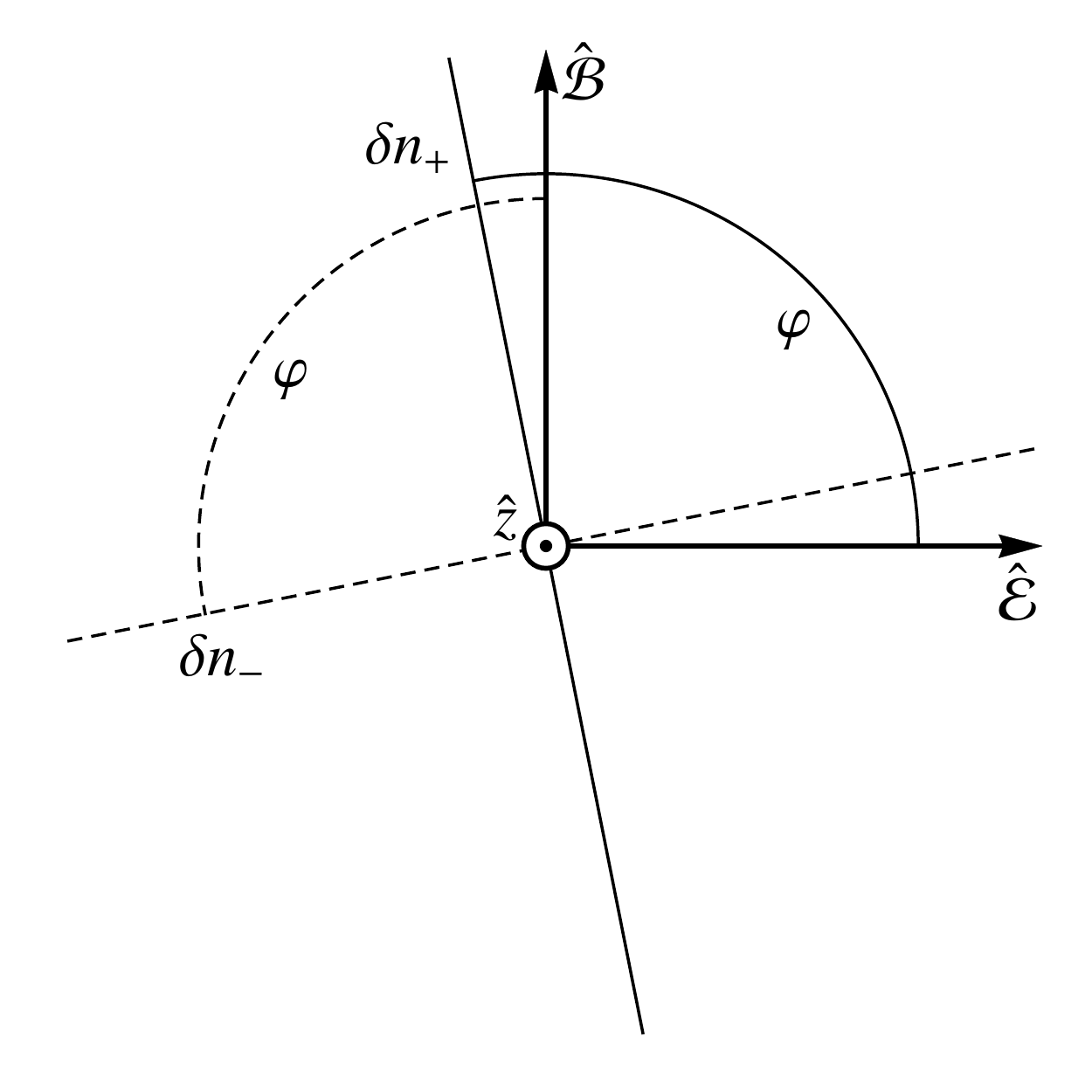}
\caption{Eigenpolarizations for the electric field amplitude ${\bf e}^{(0)}$ of the probe, which propagates in the $-z$-direction (i.e., into the page).  
They lie in the $xy$-plane, rotated with respect to $\hat{\mathcal{E}}$ and $\hat{\mathcal{B}}$ by the angle $\varphi$ whose value is determined by Eqs.~(\ref{eq:pol_rotation}).  
The polarizations corresponding to refractive index changes $\delta n_{+}$ and $\delta n_{-}$ of Eq.~(\ref{eq:delta-n_constant}) are given by the left and right columns of $R\left(\varphi\right)$ of Eq.~(\ref{eq:rotation_matrix}) and shown here in solid and dashed line, respectively.  
The left and right panels differ by a rotation of $\pi/2$, which (in effect) leaves the eigenpolarization directions invariant but switches the associated refractive indices.  
In the EH model, we find $\varphi = \pi/2$, so the $\delta n_{+}$ polarization is aligned with $\hat{\mathcal{B}}$ while the $\delta n_{-}$ polarization is aligned with $\hat{\mathcal{E}}$.
\label{fig:polarization}}
\end{figure}

The eigenvectors of Eq.~(\ref{eq:eigenproblem_constant}) give the two eigenpolarizations of ${\bf e}^{(0)}$.  Since the matrix on the left-hand side is real and symmetric, the eigenvectors are necessarily real and (when normalized) form the columns of a two-dimensional rotation matrix
\begin{equation}
R\left(\varphi\right) = \left[ \begin{array}{cc} {\rm cos} \varphi & - {\rm sin} \varphi \\ {\rm sin} \varphi & {\rm cos} \varphi \end{array} \right] \, .
\label{eq:rotation_matrix}
\end{equation}
The parameter $\varphi$ is simply the angle through which the eigenvectors are rotated with respect to the basis $\left\{ \hat{\mathcal{E}} \,,\, \hat{\mathcal{B}} \right\}$ (see Fig.~\ref{fig:polarization}), and is defined up to a multiple of $\pi$ since a half-rotation simply flips the signs of the eigenvectors without changing their orientation.  $\varphi$ can thus be chosen to lie in the half-open interval $\left(-\pi/2 \,,\, \pi/2\right]$, and the matrix on the left-hand side of Eq.~(\ref{eq:eigenproblem_constant}) can be written as $R(\varphi) D R^{-1}(\varphi)$, where $D$ is a diagonal matrix whose entries are the eigenvalues of Eq.~(\ref{eq:delta_constant}).  We take $\delta_{+}$ to be the first diagonal component of $D$, so that the left column of $R\left(\varphi\right)$ gives the polarization with the larger refractive index change, $\delta n_{+}$.  A direct calculation shows that $\varphi$ must satisfy
\begin{equation}
\delta_{1}-\delta_{2} = \left(\delta_{+}-\delta_{-}\right) {\rm cos}\left(2\varphi\right) \,, \qquad 2 \delta_{3} = \left(\delta_{+}-\delta_{-}\right) {\rm sin}\left(2\varphi\right) \,.
\label{eq:pol_rotation}
\end{equation}
There are three special cases.  First, when $\delta_{3} = 0$ and $\delta_{1}-\delta_{2} \neq 0$ (as in the EH model), we have $\varphi = 0$ or $\pi/2$ (depending on the sign of $\delta_{1}-\delta_{2}$) and the polarizations are aligned with the vectors $\hat{\mathcal{E}}$ and $\hat{\mathcal{B}}$. 
Second, when $\delta_{1}-\delta_{2} = 0$ and $\delta_{3} \neq 0$, we have $\varphi = \pm \pi/4$, so the polarizations are at $45^{\degree}$ to $\hat{\mathcal{E}}$ and $\hat{\mathcal{B}}$.
Third, when both $\delta_{3} = 0$ and $\delta_{1}-\delta_{2}=0$ (as in the BI model), $\varphi$ is undefined, but this is not a problem as there is an absence of birefringence (and hence also of well-defined eigenpolarizations) in this case.

\section{Optical Kerr effect
\label{sec:optical_Kerr}}

In this section we turn to the main focus of this paper: the refractive index change of vacuum engendered by an intense propagating wave or ``pump'', which provides the strong background fields described in Sec.~\ref{sec:effective_medium}.  We derive the dependence on the tilt angle (between the propagation directions of pump and probe), as well as the effect of elliptical polarization of the pump.  Finally, we express the results in terms of the wave intensity (rather than the field strength) in order to extract the equivalent of the nonlinear 
index for vacuum by analogy with Eq.~(\ref{eq:Kerr_index_defn}). %

\subsection{Fields of a monochromatic pump wave}

Let us consider then the fields of a propagating beam, which 
we assume %
can be approximated as monochromatic over spacetime regions much larger than the wavelength and longer than the period of the probe.  We may again use Eqs.~(\ref{eq:eigenvalue_eqn_2x2}) and~(\ref{eq:vectors_defn}), though now the overbars extracting the ``slowly-varying'' components of the outer products in Eq.~(\ref{eq:eigenvalue_eqn_2x2}) will come into play.  We write the pump fields in the form
\begin{eqnarray}
{\bf E}_{0} & = & \frac{1}{2} \, {\bf E}_{0}^{(0)} \, e^{i {\bf k}_{0} \cdot {\bf r} - i \omega_{0} t} + {\rm c.c.} \,, \nonumber \\
{\bf B}_{0} & = & \frac{1}{2} \, {\bf B}_{0}^{(0)} \, e^{i {\bf k}_{0} \cdot {\bf r} - i \omega_{0} t} + {\rm c.c.} \,,
\label{eq:pump_fields}
\end{eqnarray}
where, in order to satisfy the Maxwell equations~(\ref{eq:Maxwell}) with ${\bf D} = {\bf E}$ and ${\bf H} = {\bf B}$ (as a single plane wave must, having $\mathcal{F}_{0} = \mathcal{G}_{0} = 0$), we have $\omega_{0} = c k_{0}$ (where $k_{0} = \left|{\bf k}_{0}\right|$), and
\begin{equation}
{\bf B}_{0}^{(0)} = \frac{{\bf k}_{0}}{k_{0}} \times {\bf E}_{0}^{(0)} \,.
\label{eq:pw_BE}
\end{equation}
Note that we do not specify the direction of ${\bf k}_{0}$, whose orientation with respect to the probe wave vector ${\bf k} = -k \hat{z}$ is taken to be arbitrary.

The next step is to work out the vectors $\mathcal{E}$ and $\mathcal{B}$ entering the matrix in Eq.~(\ref{eq:eigenvalue_eqn_2x2}), before application of the overbars.  These will be oscillatory just as ${\bf E}$ and ${\bf B}$ are, and we may write: 
\begin{eqnarray}
\mathcal{E} & = & \frac{1}{2} \, \mathcal{E}^{(0)} \, e^{i {\bf k}_{0} \cdot {\bf r} - i \omega_{0} t} + {\rm c.c.} \,, \nonumber \\
\mathcal{B} & = & \frac{1}{2} \, \mathcal{B}^{(0)} \, e^{i {\bf k}_{0} \cdot {\bf r} - i \omega_{0} t} + {\rm c.c.} \,.
\label{eq:vectors_oscillatory}
\end{eqnarray}
where the amplitude vectors are given by
\begin{alignat}{4}
\mathcal{E}^{(0)} & = &\,\, {\bf E}^{(0)}_{0, \perp} - \hat{z} \times {\bf B}^{(0)}_{0, \perp} & = &\,\, -\hat{z} \times \left[ \left( \hat{z} + \frac{{\bf k}_{0}}{k_{0}} \right) \times {\bf E}_{0}^{(0)} \right] \,, \nonumber \\
\mathcal{B}^{(0)} & = &\,\, {\bf B}^{(0)}_{0, \perp} + \hat{z} \times {\bf E}^{(0)}_{0, \perp} & = &\,\, -\hat{z} \times \left[ \left( \hat{z} + \frac{{\bf k}_{0}}{k_{0}} \right) \times {\bf B}_{0}^{(0)} \right] \,,
\label{eq:amplitude_vectors_oscillatory}
\end{alignat}
Here, we have used Eq.~(\ref{eq:pw_BE}), as well as standard identities concerning two successive applications of the cross product.  These vectors evidently lie in the $xy$-plane, and by construction we again have $\mathcal{B}^{(0)} = \hat{z} \times \mathcal{E}^{(0)}$ and $\mathcal{E}^{(0)} = -\hat{z} \times \mathcal{B}^{(0)}$.  
After some further algebra, the squared magnitude of $\mathcal{E}^{(0)}$ (and hence also of $\mathcal{B}^{(0)}$) can be shown to be
\begin{eqnarray}
\mathcal{E}^{(0)\star} \cdot \mathcal{E}^{(0)} & = & \left(1 + \frac{\hat{z} \cdot {\bf k}_{0}}{k_{0}} \right)^{2} \left| {\bf E}_{0}^{(0)} \right|^{2} \nonumber \\
& = & \left(1 + {\rm cos}\theta\right)^{2} \left| {\bf E}_{0}^{(0)} \right|^{2} \nonumber \\
& = & 4 \, {\rm cos}^{4} \frac{\theta}{2} \left| {\bf E}_{0}^{(0)} \right|^{2} \,.
\label{eq:real_vectors_magnitude}
\end{eqnarray}
Here, we have introduced the {\it tilt angle} $\theta$ between the wave vectors of the pump and probe (illustrated in Fig.~\ref{fig:waves}).  This is defined to be zero when the pump and probe are exactly counter-propagating and $\pm\pi$ when they are exactly co-propagating.  
Note that the magnitude of $\mathcal{E}^{(0)}$ vanishes in the latter case, this being consistent with 
the observation made after Eqs.~(\ref{eq:constitutive_NL_2}) that there is no SPM in vacuum; on the other hand, it is maximal when the pump and probe are exactly counter-propagating.

\subsection{Accounting for elliptical polarization of the pump}

Inserting Eqs.~(\ref{eq:vectors_oscillatory}) into Eq.~(\ref{eq:eigenvalue_eqn_2x2}), and implementing the overbars by dropping all rapidly oscillating terms, we are led to the following eigenproblem:
\begin{equation}
\frac{1}{4} \, \mathrm{Re}\left\{ \delta_{1} \, \mathcal{E}^{(0) \star} \mathcal{E}^{(0) {\rm T}} + \delta_{2} \, \mathcal{B}^{(0) \star} \mathcal{B}^{(0) {\rm T}} + \delta_{3} \left( \mathcal{E}^{(0) \star} \mathcal{B}^{(0) {\rm T}} + \mathcal{B}^{(0) \star} \mathcal{E}^{(0) {\rm T}} \right) \right\} {\bf e}^{(0)} = \delta n \, {\bf e}^{(0)} \,.
\label{eq:eigenvalue_eqn_2x2_oscillatory}
\end{equation}
Compared to the case of 
constant background fields studied in Sec.~\ref{sub:constantfields}, 
we have here a complication in that 
the field amplitudes ${\bf E}^{(0)}_{0}$ and ${\bf B}^{(0)}_{0}$, and by extension the vectors $\mathcal{E}^{(0)}$ and $\mathcal{B}^{(0)}$, are generally complex.   
It is thus no longer convenient to use an orthonormal basis aligned with $\mathcal{E}^{(0)}$ and $\mathcal{B}^{(0)}$, since the matrix on the left-hand side of Eq.~(\ref{eq:eigenvalue_eqn_2x2_oscillatory}) also depends on $\mathcal{E}^{(0) \star}$ and $\mathcal{B}^{(0) \star}$, and in general $\mathcal{E}^{(0) \star} \neq \mathcal{E}^{(0)}$ and $\mathcal{B}^{(0) \star} \neq \mathcal{B}^{(0)}$.  
The issue is not with overall phases -- it is clear that the matrix in question is invariant under equal overall phase rotations of $\mathcal{E}^{(0)}$ and $\mathcal{B}^{(0)}$ -- but with relative phases between the components of these vectors. 
Such relative phases are directly related to the degree of elliptical polarization of the pump wave.  
If we were to restrict our attention to a linearly polarized pump, there would be no such relative phase, $\mathcal{E}^{(0)}$ and $\mathcal{B}^{(0)}$ could be defined to be real, and the eigenproblem would be equivalent to that of Eq.~(\ref{eq:eigenvalue_eqn_2x2_constantfields}) (except for an overall factor of $1/2$ stemming from the average over rapidly oscillating terms).
We can thus expect to recover almost the same results as those of Sec.~\ref{sub:constantfields} when the pump is linearly polarized.
However, with a proper treatment 
of the complex vectors entering the left-hand side of Eq.~(\ref{eq:eigenvalue_eqn_2x2_oscillatory}), the effects of a general elliptical polarization can be fully included in the analysis.

\begin{figure}
\includegraphics[width=0.45\columnwidth]{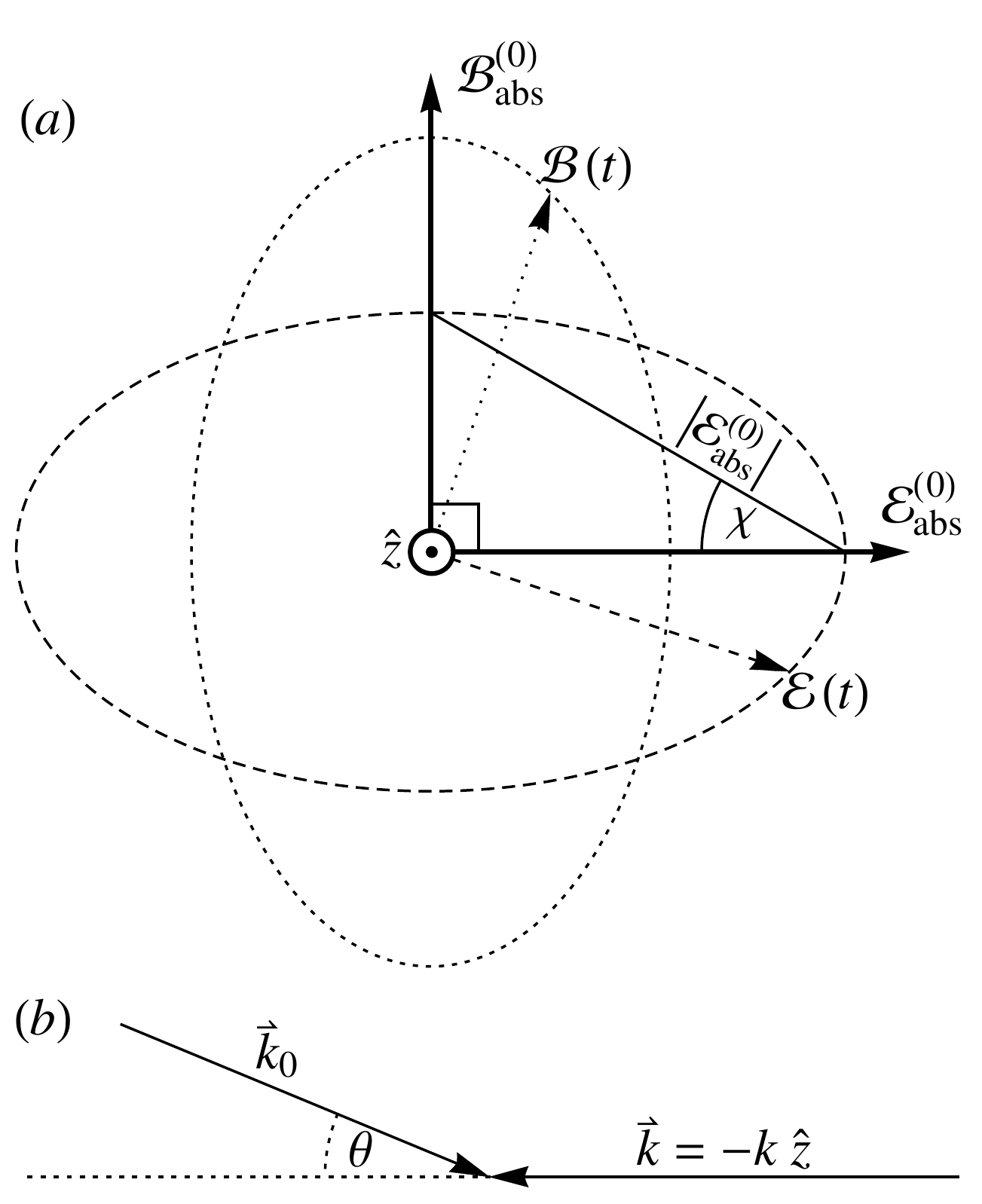}
\caption{Relative orientations for a monochromatic pump wave.  In $(a)$ are shown the vectors $\mathcal{E}(t)$ and $\mathcal{B}(t)$ (in dashed and dotted line, respectively) in the $xy$-plane, with $\hat{z}$ pointing out of the page and the wave vector of the probe pointing into the page.  These oscillate in time so as to trace out ellipses, 
rotated with respect to each other by $90 \degree$. 
The constant real vectors $\mathcal{E}_{\rm abs}^{(0)}$ and $\mathcal{B}_{\rm abs}^{(0)}$ are defined to lie along the semi-major axes of their respective 
ellipses, with magnitude equal to the length of the hypotenuse of the right-angled triangle shown.  The ellipticity angle $\chi$ is formed by the same right-angled triangle, though its sign is determined by the sense of rotation of $\mathcal{E}(t)$ and $\mathcal{B}(t)$. 
In $(b)$ are shown the wave vectors of the pump and probe waves forming the tilt angle $\theta$, 
equal to $0$ for exactly counter-propagating waves and $\pm\pi$ for exactly co-propagating waves.
\label{fig:waves}}
\end{figure}

Up to an overall phase, 
the complex field amplitudes ${\bf E}^{(0)}_{0}$ and ${\bf B}^{(0)}_{0}$ can always be written as 
a sum of two orthogonal real vectors, one (which for definiteness we 
take to be the larger in magnitude of the two) with a purely real coefficient, 
the other with a purely imaginary coefficient.  
These vectors necessarily lie in the plane perpendicular to the wave vector ${\bf k}_{0}$.  They describe the principle directions of the polarization, i.e., the major and minor axes of the ellipse formed by the oscillating electric or magnetic field (see Fig.~\ref{fig:waves}).  The field lies along the minor axis exactly a quarter of a period after it lies along the major axis, and so the complex components of the field along these two directions appear with a relative phase of $\pi/2$.  We may thus write the field amplitudes using the real 
vectors ${\bf E}^{(0)}_{\rm abs}$ and ${\bf B}^{(0)}_{\rm abs}$, 
and again utilizing the fact that ${\bf B}^{(0)}_{0} = \left({\bf k}_{0}/k_{0}\right) \times {\bf E}^{(0)}_{0}$:
\begin{eqnarray}
{\bf E}^{(0)}_{0} & = & e^{i \phi_{0}} \left( {\rm cos}\chi \, {\bf E}^{(0)}_{\rm abs} + i \, {\rm sin}\chi \, {\bf B}^{(0)}_{\rm abs} \right) \,, \nonumber \\
{\bf B}^{(0)}_{0} & = & e^{i \phi_{0}} \left( {\rm cos}\chi \, {\bf B}^{(0)}_{\rm abs} - i \, {\rm sin}\chi \, {\bf E}^{(0)}_{\rm abs} \right) \,.
\label{eq:real_vectors_normalized}
\end{eqnarray}
where ${\bf B}^{(0)}_{\rm abs} = \left({\bf k}_{0}/k_{0}\right) \times {\bf E}^{(0)}_{\rm abs}$ and ${\bf E}^{(0)}_{\rm abs} = -\left({\bf k}_{0}/k_{0}\right) \times {\bf B}^{(0)}_{\rm abs}$.
With this writing, 
${\bf E}_{0}^{(0)}$ and 
${\bf E}_{\rm abs}^{(0)}$ 
have exactly the same magnitude. 
The angle $\chi \in \left[ -\pi/4 \,,\, \pi/4 \right]$
is the so-called {\it ellipticity angle} of the polarization ellipse (also illustrated in Fig.~\ref{fig:waves}).  When $\chi = 0$, the wave is linearly polarized; when $\chi = \pm \pi/4$, it is circularly polarized, in which case the directions of ${\bf E}^{(0)}_{\rm abs}$ and ${\bf B}^{(0)}_{\rm abs}$ in the 2D plane can be chosen arbitrarily.
$\chi$ is 
defined to be positive when the sense of rotation of the fields is from ${\bf E}^{(0)}_{\rm abs}$ towards ${\bf B}^{(0)}_{\rm abs}$, corresponding to 
right-handed polarization of the pump. 

By a straightforward application of Eqs.~(\ref{eq:amplitude_vectors_oscillatory}), we may now write
\begin{eqnarray}
\mathcal{E}^{(0)} & = & e^{i \phi_{0}} \, \left( {\rm cos} \chi \, \mathcal{E}_{\rm abs}^{(0)} + i \, {\rm sin} \chi \, \mathcal{B}_{\rm abs}^{(0)} \right) \,, \nonumber \\
\mathcal{B}^{(0)} & = & e^{i \phi_{0}} \, \left( {\rm cos} \chi \, \mathcal{B}_{\rm abs}^{(0)} - i \, {\rm sin} \chi \, \mathcal{E}_{\rm abs}^{(0)} \right) \,,
\label{eq:vectors_xy_decomp}
\end{eqnarray}
where we have defined
\begin{eqnarray}
\mathcal{E}_{\rm abs}^{(0)} & = & -\hat{z} \times \left[ \left( \hat{z} + \frac{{\bf k}_{0}}{k_{0}} \right) \times {\bf E}_{\rm abs}^{(0)} \right] \,, \nonumber \\
\mathcal{B}_{\rm abs}^{(0)} & = & -\hat{z} \times \left[ \left( \hat{z} + \frac{{\bf k}_{0}}{k_{0}} \right) \times {\bf B}_{\rm abs}^{(0)} \right] \,.
\label{eq:vectors_real_amplitudes}
\end{eqnarray}
It is again fairly straightforward to show that $\mathcal{B}_{\rm abs}^{(0)} = \hat{z} \times \mathcal{E}_{\rm abs}^{(0)}$ and $\mathcal{E}_{\rm abs}^{(0)} = -\hat{z} \times \mathcal{B}_{\rm abs}^{(0)}$, so that these 
are orthogonal and have equal magnitude.  
Therefore, analogously to Eqs.~(\ref{eq:real_vectors_normalized}), Eqs.~(\ref{eq:vectors_xy_decomp}) define an ellipse of ellipticity angle $\chi$ in the $xy$-plane. 
It is particularly interesting that the ellipticity angle remains $\chi$ when passing to $\mathcal{E}^{(0)}$ and $\mathcal{B}^{(0)}$, since these 
lie in the $xy$-plane rather than the plane containing ${\bf E}_{\rm abs}^{(0)}$ and ${\bf B}_{\rm abs}^{(0)}$. 
One might naively have expected the shape of the ellipse to depend on the orientation of ${\bf k}_{0}$ with respect to $\hat{z}$ 
(as the orthogonal projections of the ellipses traced out by the electric and magnetic fields certainly do depend on the angle they are viewed from).  Remarkably, however, it turns out that $\mathcal{E}^{(0)}$ and $\mathcal{B}^{(0)}$ combine the electric and magnetic fields of the pump wave in just the right way so that $\chi$ is invariant with respect to the relative orientation of pump and probe.
Only the magnitude of $\mathcal{E}^{(0)}$ (and hence of $\mathcal{E}_{\rm abs}^{(0)}$) varies with the orientation, %
as shown by Eq.~(\ref{eq:real_vectors_magnitude}).

\subsection{Nonlinear index of vacuum
\label{sub:NLindex}}

Finally, 
Eqs.~(\ref{eq:vectors_xy_decomp}) are inserted into 
Eq.~(\ref{eq:eigenvalue_eqn_2x2_oscillatory}), upon which 
it becomes the following eigenproblem:
\begin{equation}
\frac{1}{4} \left[ \delta_{1}^{\prime} \, \mathcal{E}_{\rm abs}^{(0)} \mathcal{E}_{\rm abs}^{(0) {\rm T}} + \delta_{2}^{\prime} \, \mathcal{B}_{\rm abs}^{(0)} \mathcal{B}_{\rm abs}^{(0) {\rm T}} + \delta_{3}^{\prime} \, \left(\mathcal{E}_{\rm abs}^{(0)} \mathcal{B}_{\rm abs}^{(0) {\rm T}} + \mathcal{B}_{\rm abs}^{(0)} \mathcal{E}_{\rm abs}^{(0) {\rm T}} \right) \right] {\bf e}^{(0)} = \delta n \, {\bf e}^{(0)} \,,
\label{eq:eigenvalue_eqn_2x2_osc_real}
\end{equation}
where 
\begin{eqnarray}
\delta_{1}^{\prime} + \delta_{2}^{\prime} & = & \delta_{1} + \delta_{2} \,, \nonumber \\
\delta_{1}^{\prime} - \delta_{2}^{\prime} & = & \left( \delta_{1} - \delta_{2} \right) {\rm cos}\left(2\chi\right) \,, \nonumber \\
\delta_{3}^{\prime} & = & \delta_{3} \, {\rm cos}\left(2\chi\right) \,.
\label{eq:delta_pol}
\end{eqnarray}
This is now in a form completely analogous to Eq.~(\ref{eq:eigenvalue_eqn_2x2_constantfields}), 
with the vectors $\mathcal{E}_{\rm abs}^{(0)}$ and $\mathcal{B}_{\rm abs}^{(0)}$ being equal in magnitude and 
orthogonal to each other. 
They can thus be used to define basis vectors in the $xy$-plane, 
$\hat{\mathcal{E}} = \mathcal{E}_{\rm abs}^{(0)}/\left| \mathcal{E}_{\rm abs}^{(0)} \right|$ and $\hat{\mathcal{B}} = \mathcal{B}_{\rm abs}^{(0)}/\left| \mathcal{B}_{\rm abs}^{(0)} \right|$, so that the vectors $\left\{\hat{\mathcal{E}} \,,\, \hat{\mathcal{B}} \,,\, \hat{z}\right\}$ form a right-handed orthonormal basis.  
Restricting our attention to the $xy$-plane in the basis $\left\{ \hat{\mathcal{E}} \,,\, \hat{\mathcal{B}} \right\}$, 
Eq.~(\ref{eq:eigenvalue_eqn_2x2_osc_real}) can be written in matrix form as follows:
\begin{equation}
\Bigg[ \begin{array}{cc} \delta_{1}^{\prime} & \delta_{3}^{\prime} \\ \delta_{3}^{\prime} & \delta_{2}^{\prime} \end{array} \Bigg] \left[ \begin{array}{c} e^{(0)}_{\mathcal{E}} \\ e^{(0)}_{\mathcal{B}} \end{array} \right] = \frac{\delta n}{\frac{1}{4} \left|\mathcal{E}_{\rm abs}^{(0)}\right|^{2}} \left[ \begin{array}{c} e^{(0)}_{\mathcal{E}} \\ e^{(0)}_{\mathcal{B}} \end{array} \right] \,.
\label{eq:opticalKerr_matrix_eqn}
\end{equation}
The eigenvalues of the matrix on the left-hand side of this equation are readily found, and give the possible values of $\delta n$.  Using Eq.~(\ref{eq:real_vectors_magnitude}) and the fact that $\mathcal{E}^{(0)}$ and $\mathcal{E}_{\rm abs}^{(0)}$ are defined to have the same magnitude, 
we have
\begin{equation}
\delta n_{\pm} = \delta_{\pm}^{\prime} \, {\rm cos}^{4} \frac{\theta}{2} \, \left|{\bf E}_{0}^{(0)}\right|^{2} \,,
\label{eq:delta-n_optical}
\end{equation}
where
\begin{eqnarray}
\delta_{\pm}^{\prime} & = & \frac{1}{2} \left( \delta_{1}^{\prime} + \delta_{2}^{\prime} \pm \sqrt{ \left(\delta_{1}^{\prime}-\delta_{2}^{\prime}\right)^{2} + \left(2 \delta_{3}^{\prime}\right)^{2} } \right) \nonumber \\
& = & \frac{1}{2} \left( \delta_{1} + \delta_{2} \pm {\rm cos}\left(2\chi\right) \, \sqrt{ \left(\delta_{1}-\delta_{2}\right)^{2} + \left(2 \delta_{3}\right)^{2} } \right) \,.
\label{eq:delta_optical}
\end{eqnarray}
We thus see that the ellipticity angle $\chi$ directly affects the strength of the birefringence, which vanishes completely in the case of circular polarization.  
Moreover, Eqs.~(\ref{eq:pol_rotation}) still hold (being simply multiplied by an overall factor of $\mathrm{cos}\left(2\chi\right)$), so that the rotation angle $\varphi$ of the eigenpolarizations with respect to the $\left\{\hat{\mathcal{E}} \,,\, \hat{\mathcal{B}}\right\}$ basis is independent of $\chi$.
Note that the probe eigenstates 
are linearly polarized no matter the polarization state of the pump, a direct result of the fact that the matrix on the left-hand side of Eq.~(\ref{eq:eigenvalue_eqn_2x2_oscillatory}) is real and symmetric.  (In 
Sec.~\ref{sec:axions} 
we shall examine a model where this is no longer the case, 
allowing complex eigenvectors which encode states of 
elliptical polarization.) 

We are now in a position 
to give explicit values for the corresponding 
nonlinear 
index $n_{2}$, defined (by analogy with Eq.~(\ref{eq:Kerr_index_defn})) such that the refractive index change $\delta n = n_{2} \, I$, where $I$ is the intensity of the pump wave.  
For definiteness and simplicity, we consider pump and probe to be exactly counter-propagating ($\theta = 0$), this being the optimal geometry~\footnote{In the context of the EH model, Ref.~\cite{Aleksandrov-et-al-1985} seems to have been the first to study this particular geometry.  Ref.~\cite{Ferrando-et-al-2007} studied 
a suboptimal geometry in which the waves have parallel polarizations and a tilt angle of $90\degree$, yielding an overall reduction factor of $7$ in the value of $n_{2}$ with respect to the optimal value, $n_{2,\perp}^{\rm (EH)}$ in Eqs.~(\ref{eq:n2-EH}). \label{fn:geometry}} according to Eq.~(\ref{eq:delta-n_optical}). 
The energy density of the pump (after averaging over rapidly oscillating terms) is $\left|{\bf E}_{0}^{(0)}\right|^{2}/2$, and its intensity is found upon multiplication 
by $c$.  
The corresponding values of $n_{2}$ are then simply $\delta_{\pm}^{\prime} \times 2/c$. 
In the EH model 
we have, for a linearly polarized pump beam, 
\begin{equation}
n_{2, \parallel}^{\rm (EH)} \approx 0.888 \times 10^{-33} \, {\rm cm}^{2} / {\rm W} \,, \qquad n_{2, \perp}^{\rm (EH)} \approx 1.555 \times 10^{-33} \, {\rm cm}^{2} / {\rm W} \,,
\label{eq:n2-EH}
\end{equation}
where the subscripts `$\parallel$' and `$\perp$' refer to the probe and pump fields being equally and orthogonally polarized, respectively. 
On the other hand, for a circularly polarized pump, 
$n_{2}$ no longer depends on the polarization of the probe 
and is simply the arithmetic mean of the two values given above: 
\begin{equation}
n_{2, {\rm circ}}^{\rm (EH)} \approx 1.222 \times 10^{-33} \, {\rm cm}^{2} / {\rm W} \,.
\end{equation}
It can thus be seen that the EH prediction for the nonlinear index of vacuum, though it depends on pump polarization and tilt angle, is on the order of $10^{-33} \, {\rm cm}^{2}/{\rm W}$.  

In the BI model instead, $\delta_{\pm}^{\prime}$ 
of Eqs.~(\ref{eq:delta_optical}) are both equal to $\delta^{\rm (BI)}$ and independent of the ellipticity angle $\chi$. 
We thus have simply
\begin{equation}
n_{2}^{\rm (BI)} = \frac{2}{c \, b^{2}} \,,
\end{equation}
where $b^{2}$ is the square of the critical field parametrizing the BI model (expressed in units of energy density). 
Using the value of $\delta^{\rm (BI)}$ in Eq.~(\ref{eq:BI-coeff}), with $b$ being the field at the classical radius of the electron, this gives
\begin{equation}
n_{2}^{\rm (BI)} \approx 0.229 \times 10^{-33} \, {\rm cm}^{2} / {\rm W} \,.
\end{equation}





\section{A dispersive model: coupling to axions
\label{sec:axions}}

Experiments in NLED have been considered as potentially enabling the detection of the axion~\cite{Sikivie-1983,Maiani-Petronzio-Zavattini-1986,Gasperini-1987,vanBibber-et-al-1987,Raffelt-Stodolsky-1988}, a hypothetical particle introduced as a possible explanation for strong CP invariance in quantum chromodynamics~\cite{Peccei-Quinn-1977-PRL,Peccei-Quinn-1977-PRD,Wilczek-1978,Weinberg-1978}, and which has been proposed as a candidate for dark matter~\cite{Arias-et-al-2012}.
As far as electromagnetism is concerned, the axion field couples directly to $\mathcal{G} = {\bf E} \cdot {\bf B}$, and will thus contribute to the effective photon/photon interaction in NLED.
However, there are compelling astrophysical~\cite{Raffelt-2008} and cosmological~\cite{Marsh-2016} reasons to consider an axion mass which is significantly smaller than $1\,{\rm eV}$, the energy scale of an optical photon.
In this case, the Compton wavelength of the axion is long compared to the typical photon wavelength, and the assumption of purely local effective photon/photon interactions made in Sec.~\ref{sec:preliminaries} is explicitly broken.
Coupling to the axion field is thus not only of potential experimental relevance (though the question of experimental feasibility is beyond the scope of this paper), but is also of theoretical interest as it will lead to a dispersive model of NLED (as was recently illustrated in~\cite{Baldenegro-et-al-2019}).

In this section, starting from the electromagnetic Lagrangian of Eq.~(\ref{eq:Lagrangian}), we couple the electromagnetic field to an axion field.
The analysis of previous sections is carried through in a similar manner; the post-Maxwellian parameters and the polarization of the pump are left unspecified, thereby generalizing the results of~\cite{Baldenegro-et-al-2019} (which considered the EH model and a linearly polarized pump).
We shall assume that the various plane waves are infinite in extent and duration.  
This allows us to neglect 
retardation effects due to the non-instantaneous nature of the axion response, including photon-axion oscillations~\cite{Raffelt-Stodolsky-1988}. 
Instead, we here focus solely on the refractive index change induced by the axion coupling.

\subsection{Lagrangian and constitutive relations}

The total effective Lagrangian, including the coupling to axions, may be written
\begin{equation}
\mathcal{L} = \mathcal{L}_{\rm EM} + \mathcal{L}_{\rm ax} + \mathcal{L}_{\rm int} \,.
\end{equation}
Here, 
$\mathcal{L}_{\rm EM}$ can be considered as the ``local part'' of the 
effective Lagrangian containing 
the electromagnetic fields alone, 
and is just that used in previous sections (and given in Eq.~(\ref{eq:Lagrangian})).  
The next term is the Lagrangian of the free axion field: 
\begin{equation}
\mathcal{L}_{\rm ax} = \frac{1}{2}\left(\partial_{c t} \phi\right)^{2} - \frac{1}{2} \left( \nabla \phi \right)^{2} - \frac{1}{2} k_{\rm ax}^{2} \phi^{2} \,,
\end{equation}
where $k_{\rm ax} = m_{\rm ax} c/\hbar$ is the wave vector associated to the Compton wavelength of the axion.  Finally, the interaction between the axion and electromagnetic fields is described by 
\begin{equation}
\mathcal{L}_{\rm int} = - \eta \, \phi \, \mathcal{G} \,,
\end{equation}
where the inverse square of the coupling constant, $\eta^{-2}$, has dimensions of energy per unit length.

Employing the separation into background and probe fields described in Sec.~\ref{sub:linearization}, and using a similar decomposition for the axion field $\phi = \phi_{0} + \delta\phi$, we may write the total Lagrangian as
\begin{eqnarray}
\mathcal{L} & = & \mathcal{L}_{\rm EM}\left({\bf E}_{0} + {\bf e} \,,\, {\bf B}_{0} + {\bf b}\right) + \mathcal{L}_{\rm ax}\left(\phi_{0}+\delta\phi\right) + \mathcal{L}_{\rm int}\left({\bf E}_{0}+{\bf e}\,,\,{\bf B}_{0}+{\bf b}\,,\,\phi+\delta\phi\right) \nonumber \\
& \approx & \mathcal{L}_{\rm 0} + \mathcal{L}_{\rm probe}
\end{eqnarray}
where
\begin{equation}
\mathcal{L}_{\rm 0} = \mathcal{L}_{\rm EM}\left({\bf E}_{0} \,,\, {\bf B}_{0} \right) + \mathcal{L}_{\rm ax}\left(\phi_{0}\right) + \mathcal{L}_{\rm int}\left({\bf E}_{0} \,,\, {\bf B}_{0} \,,\, \phi_{0}\right)
\end{equation}
is the Lagrangian associated with the background alone
and, keeping only terms which are quadratic in the weak 
fields of 
the probe,
\begin{equation}
\mathcal{L}_{\rm probe} = \mathcal{L}_{\rm EM, probe}\left({\bf E}_{0}\,,\,{\bf B}_{0} \,;\, {\bf e}\,,\,{\bf b}\right) + \mathcal{L}_{\rm ax}\left(\delta\phi\right) - \eta \left( \phi_{0} \, {\bf e} \cdot {\bf b} + \delta\phi \, {\bf E}_{0} \cdot {\bf b} + \delta\phi \, {\bf B}_{0} \cdot {\bf e} \right) \,.
\end{equation}
$\mathcal{L}_{\rm EM, probe}$ is exactly the probe Lagrangian derived in Sec.~\ref{sub:linearization}, while $\mathcal{L}_{\rm ax}$ (which is already purely quadratic) is again just the Lagrangian for a free axion field.  The terms proportional to $\eta$ describe the interplay between the electromagnetic and axion fields associated with the passage of the probe wave.  
Since we deal with a restricted class of background configurations, we can simplify this term further. 
Noting that we deal either with plane wave background fields, which satisfy $\mathcal{G}_{0} = 0$ and hence $\phi_{0}=0$, or with constant background fields, in which case the term in ${\bf e} \cdot {\bf b}$ is a total derivative when expressed in terms of the vector potential (see the discussion following Eqs.~(\ref{eq:scalars}), including footnote~\ref{fn:total_derivative}), the term $\phi_{0}\,{\bf e}\cdot{\bf b}$ can be removed from the Lagrangian with no effect on the field equations.  Therefore, we consider only the terms proportional to $\delta\phi$ as far as the coupling to the axion field is concerned.  

Applying definitions~(\ref{eq:constitutive}) to the probe fields, we have
\begin{eqnarray}
{\bf d} & = & {\bf d}_{\rm EM} - \eta \, \delta\phi \, {\bf B}_{0} \,, \nonumber \\
{\bf h} & = & {\bf h}_{\rm EM} + \eta \, \delta\phi \, {\bf E}_{0} \,,
\label{eq:constitutive_axion}
\end{eqnarray}
where ${\bf d}_{\rm EM}$ and ${\bf h}_{\rm EM}$ are due to $\mathcal{L}_{\rm EM, probe}$ alone and are already defined in Eqs.~(\ref{eq:probe_constitutive_1}).  
Our aim is to 
subject these relations to the same treatment as in Sec.~\ref{sec:effective_medium}; in particular, to find the new forms of Eqs.~(\ref{eq:probe_constitutive_pw}) relating the amplitudes ${\bf d}^{(0)}$ and ${\bf h}^{(0)}$ directly to ${\bf e}^{(0)}$, so that a homogeneous linear equation analogous to Eq.~(\ref{eq:eigenvalue_eqn_3x3}) is obtained. 
Our first task, then, is to determine the axion field $\delta \phi$ generated by a probe wave of amplitude ${\bf e}^{(0)}$.

\subsection{Response of axion field to passage of probe wave} 

The response of $\delta\phi$ to the presence of electromagnetic fields is 
determined by the following equation of motion:
\begin{equation}
\left[ \partial_{c t}^{2} - \nabla^{2} + k_{\rm ax}^{2} \right] \delta\phi = - \eta \, \delta\mathcal{G} \,,
\end{equation}
where we have defined $\delta\mathcal{G} = {\bf E}_{0} \cdot {\bf b} + {\bf B}_{0} \cdot {\bf e}$.
The probe-induced $\delta\mathcal{G}$ thus acts as a source for $\delta\phi$, with a simple relationship between their Fourier components:
\begin{equation}
\delta\phi_{\omega^{\prime}, k^{\prime}} = \eta \, \frac{ \delta\mathcal{G}_{\omega^{\prime}, k^{\prime}}}{\left(\omega^{\prime}/c\right)^{2} - \left(k^{\prime}\right)^{2} - k_{\rm ax}^{2}} \,,
\label{eq:axion_Fourier}
\end{equation}
assuming of course that we are not at resonance, i.e., $\left(\omega^{\prime}/c\right)^{2} - \left(k^{\prime}\right)^{2} - k_{\rm ax}^{2} \neq 0$. 
Using Eqs.~(\ref{eq:pump_fields}) for the background/pump fields (noting that they reduce to constant fields when ${\bf k}_{0}$ and $\omega_{0}$ vanish), we have, to lowest order,
\begin{equation} 
\delta \mathcal{G} 
\approx \frac{1}{4} \left\{ \left[ {\bf B}_{0}^{(0)} + \hat{z} \times {\bf E}_{0}^{(0)} \right] e^{i {\bf k}_{0} \cdot {\bf r} - i \omega_{0} t} +  \left[ {\bf B}_{0}^{(0)} + \hat{z} \times {\bf E}_{0}^{(0)} \right]^{\star} e^{-i{\bf k}_{0} \cdot {\bf r} + i \omega_{0} t} \right\} \cdot {\bf e}^{(0)} e^{-i k z - i \omega t} + {\rm c.c.} \,,
\end{equation} 
where 
we have used Eq.~(\ref{eq:pw_requirements_a}), 
as well as the cyclic invariance of the vector triple product, to write ${\bf E}_{0}^{(0)} \cdot {\bf b}_{0}^{(0)} \approx 
\left(\hat{z}\times {\bf E}_{0}^{(0)}\right) \cdot {\bf e}^{(0)}$.  Using relation~(\ref{eq:axion_Fourier}), we can immediately write down the generated axion field:
\begin{eqnarray}
\delta \phi & \approx & \frac{\eta}{4} \, \left\{ \frac{\left[{\bf B}_{0}^{(0)} + \hat{z} \times {\bf E}_{0}^{(0)}\right] e^{i{\bf k}_{0} \cdot {\bf r} - i \omega_{0} t}}{\left(\omega_{0}+\omega\right)^{2}/c^{2} - \left({\bf k}_{0} - k \hat{z} \right)^{2} - k_{\rm ax}^{2}} + \frac{\left[{\bf B}_{0}^{(0)} + \hat{z} \times {\bf E}_{0}^{(0)}\right]^{\star} e^{-i{\bf k}_{0} \cdot {\bf r} + i \omega_{0} t}}{\left(\omega_{0}-\omega\right)^{2}/c^{2} - \left({\bf k}_{0} + k \hat{z} \right)^{2} - k_{\rm ax}^{2}} \right\} \cdot {\bf e}^{(0)} e^{-ikz - i \omega t} + {\rm c.c.} \nonumber \\
& \approx & \frac{\eta}{4} \, \left\{ \frac{\left[{\bf B}_{0}^{(0)} + \hat{z} \times {\bf E}_{0}^{(0)}\right] e^{i{\bf k}_{0} \cdot {\bf r} - i \omega_{0} t}}{4 \, {\rm cos}^{2}\frac{\theta}{2} \, \omega_{0}\omega/c^{2} - k_{\rm ax}^{2}} - \frac{\left[{\bf B}_{0}^{(0)} + \hat{z} \times {\bf E}_{0}^{(0)}\right]^{\star} e^{-i{\bf k}_{0} \cdot {\bf r} + i \omega_{0} t}}{4 \, {\rm cos}^{2}\frac{\theta}{2} \, \omega_{0}\omega/c^{2} + k_{\rm ax}^{2}} \right\} \cdot {\bf e}^{(0)} e^{-ikz - i \omega t} + {\rm c.c.} 
\label{eq:generated_axion_field}
\end{eqnarray}
In the second line, we have expanded the squares in the denominators, neglecting the variation of the refractive index here so that $k \approx \omega/c$, and used the fact that ${\bf k}_{0} \cdot \hat{z} = k_{0}\,{\rm cos}\theta$ where $\theta$ is the tilt angle between the pump and probe waves introduced in the previous section (and illustrated in Fig.~\ref{fig:waves}).

\subsection{Backreaction of axion field on probe fields}

Substituting expression~(\ref{eq:generated_axion_field}) for $\delta\phi$ back into the constitutive relations~(\ref{eq:constitutive_axion}) removes the explicit dependence on the axion field, giving ${\bf d}$ and ${\bf h}$ in terms of the probe amplitude ${\bf e}^{(0)}$ alone.  Since ${\bf d}_{\rm EM}$ and ${\bf h}_{\rm EM}$ are already given in Eqs.~(\ref{eq:probe_constitutive_pw}), we need focus here only on the additional terms $\delta{\bf d} = -\eta \, \delta\phi \, {\bf B}_{0}$ and $\delta{\bf h} = \eta \, \delta\phi \, {\bf E}_{0}$.  Since $\delta\phi$, ${\bf E}_{0}$ and ${\bf B}_{0}$ are all generally oscillatory,
the substitution 
generates rapidly oscillating terms that are far off-shell, much like when calculating ${\bf d}$ and ${\bf h}$ in 
Sec.~\ref{sec:effective_medium}. 
As there, we retain only those terms whose oscillations are synchronized with those of the probe, in which case we may write $\delta{\bf d} = {\rm Re}\left\{\delta{\bf d}^{(0)} {\rm exp}\left(-ikz - i\omega t\right)\right\}$ and $\delta{\bf h} = {\rm Re}\left\{\delta{\bf h}^{(0)} {\rm exp}\left(-ikz - i\omega t\right)\right\}$, with amplitudes $\delta{\bf d}^{(0)}$ and $\delta{\bf h}^{(0)}$ that are linearly related to ${\bf e}^{(0)}$:
\begin{equation}
\delta{\bf d}^{(0)} = \delta D_{\rm ax} \, {\bf e}^{(0)} \,, \qquad \delta{\bf h}^{(0)} = \delta H_{\rm ax} \, {\bf e}^{(0)} \,.
\label{eq:const_axion}
\end{equation}
$\delta D_{\rm ax}$ and $\delta H_{\rm ax}$ are $3 \times 3$ matrices, and direct substitution shows that
\begin{eqnarray}
\delta D_{\rm ax} & = & \frac{1}{8} \, \frac{\eta^{2}}{k_{\rm ax}^{2}} \, \left\{ \frac{{\bf B}_{0}^{(0)\star} \left[{\bf B}_{0}^{(0)} + \hat{z} \times {\bf E}_{0}^{(0)}\right]^{{\rm T}}}{1 - 4 \, {\rm cos}^{2}\frac{\theta}{2} \, \omega_{0}\omega/\omega_{\rm ax}^{2}} + \frac{{\bf B}_{0}^{(0)} \left[{\bf B}_{0}^{(0)} + \hat{z} \times {\bf E}_{0}^{(0)}\right]^{\star {\rm T}}}{1 + 4 \, {\rm cos}^{2}\frac{\theta}{2} \, \omega_{0}\omega/\omega_{\rm ax}^{2}} \right\} \,, \nonumber \\
\delta H_{\rm ax} & = & -\frac{1}{8} \, \frac{\eta^{2}}{k_{\rm ax}^{2}} \, \left\{ \frac{{\bf E}_{0}^{(0)\star} \left[{\bf B}_{0}^{(0)} + \hat{z} \times {\bf E}_{0}^{(0)}\right]^{{\rm T}}}{1 - 4 \, {\rm cos}^{2}\frac{\theta}{2} \, \omega_{0}\omega/\omega_{\rm ax}^{2}} + \frac{{\bf E}_{0}^{(0)} \left[{\bf B}_{0}^{(0)} + \hat{z} \times {\bf E}_{0}^{(0)}\right]^{\star {\rm T}}}{1 + 4 \, {\rm cos}^{2}\frac{\theta}{2} \, \omega_{0}\omega/\omega_{\rm ax}^{2}} \right\} \,,
\label{eq:deltaL_axion}
\end{eqnarray}
where $\omega_{\rm ax} = c k_{\rm ax} = m_{\rm ax} c^{2} / \hbar$.

\subsection{Axionic contribution to the refractive index}

Writing Eqs.~(\ref{eq:const_axion}) as linear in ${\bf e}^{(0)}$ is particularly convenient as it immediately allows us to combine these results with those of Eqs.~(\ref{eq:probe_constitutive_pw}), and then to obtain the modification to Eq.~(\ref{eq:eigenvalue_eqn_3x3}).  Indeed, it is straightforward to show that 
the matrix on the left-hand side of Eq.~(\ref{eq:eigenvalue_eqn_3x3}) is changed simply by addition of 
$\delta D_{\rm ax} - \Omega_{z} \, \delta H_{\rm ax}$, where (in accordance with the observations made just after Eq.~(\ref{eq:Omega_definition})) 
$\Omega_{z}$ acting on $\delta H_{\rm ax}$ replaces the ${\bf E}_{0}^{(0)}$ and ${\bf E}_{0}^{(0)\star}$ of the second of Eqs.~(\ref{eq:deltaL_axion}) by $\hat{z}\times{\bf E}^{(0)}_{0}$ and $\hat{z}\times{\bf E}_{0}^{(0)\star}$, respectively.  Furthermore, as discussed after Eq.~(\ref{eq:eigenvalue_eqn_3x3}), 
we need only consider the $xy$-projection of the matrix in Eq.~(\ref{eq:eigenvalue_eqn_3x3}) when working to first order in $\delta n$. 
The relevant combination of $\delta D_{\rm ax}$ and $\delta H_{\rm ax}$, once projected onto the $xy$-plane, depends only on the vector ${\mathcal B}^{(0)}$ of Eqs.~(\ref{eq:amplitude_vectors_oscillatory}), and not on ${\bf E}_{0}^{(0)}$ and ${\bf B}_{0}^{(0)}$ separately. 
In short, to the matrices on the left-hand sides of Eqs.~(\ref{eq:eigenvalue_eqn_2x2_oscillatory}) and~(\ref{eq:eigenvalue_eqn_2x2_osc_real}) must be added the following:
\begin{multline}
\left.\left(\delta D_{\rm ax} - \Omega_{z} \, \delta H_{\rm ax}\right)\right|_{xy} = \frac{1}{8} \, \frac{\eta^{2}}{k_{\rm ax}^{2}} \, \left\{ \frac{\mathcal{B}^{(0) \star} \mathcal{B}^{(0) {\rm T}}}{1 - 4 \, {\rm cos}^{2}\frac{\theta}{2} \, \omega_{0}\omega/\omega_{\rm ax}^{2}} + \frac{\mathcal{B}^{(0)} \mathcal{B}^{(0) \star {\rm T}}}{1 + 4 \, {\rm cos}^{2}\frac{\theta}{2} \, \omega_{0}\omega/\omega_{\rm ax}^{2}} \right\}  \\
\equiv \frac{1}{4} \left[ \Delta_{2,{\rm ax}}\left(\omega\right) \left\{ {\rm cos}^{2}\chi \, \mathcal{B}_{\rm abs}^{(0)} \mathcal{B}_{\rm abs}^{(0) {\rm T}} + {\rm sin}^{2}\chi \, \mathcal{E}_{\rm abs}^{(0)} \mathcal{E}_{\rm abs}^{(0) {\rm T}} \right\} + i \,\Delta_{3,{\rm ax}}\left(\omega\right) \, {\rm sin}\left(2\chi\right) \, \left\{\mathcal{E}_{\rm abs}^{(0)} \mathcal{B}_{\rm abs}^{(0) {\rm T}} - \mathcal{B}_{\rm abs}^{(0)} \mathcal{E}_{\rm abs}^{(0) {\rm T}} \right\} \right] \,,
\end{multline}
where in the last line we have used the decomposition of $\mathcal{B}^{(0)}$ given in Eqs.~(\ref{eq:vectors_xy_decomp}), and defined
\begin{equation}
\Delta_{2,{\rm ax}}\left(\omega\right) =  \frac{\eta^{2}}{k_{\rm ax}^{2}} \, \frac{1}{1-{\rm cos}^{4}\frac{\theta}{2} \,  \left(4\omega_{0}\omega/\omega_{\rm ax}^{2}\right)^{2}} \,, \qquad
\Delta_{3,{\rm ax}}\left(\omega\right) = \frac{\eta^{2}}{k_{\rm ax}^{2}} \, \frac{2 \, {\rm cos}^{2}\frac{\theta}{2} \,  \omega_{0}\omega/\omega_{\rm ax}^{2}}{1-{\rm cos}^{4}\frac{\theta}{2} \, \left(4\omega_{0}\omega/\omega_{\rm ax}^{2}\right)^{2}} \,.
\label{eq:axion_contributions}
\end{equation}
The eigenproblem thus becomes: 
\begin{equation}
\frac{1}{4} \left[ \delta_{1}^{\prime\prime}\left(\omega\right) \, \mathcal{E}_{\rm abs}^{(0)} \mathcal{E}_{\rm abs}^{(0) {\rm T}} + \delta_{2}^{\prime\prime}\left(\omega\right) \, \mathcal{B}_{\rm abs}^{(0)} \mathcal{B}_{\rm abs}^{(0) {\rm T}} + \delta_{3}^{\prime\prime}\left(\omega\right) \, \mathcal{E}_{\rm abs}^{(0)} \mathcal{B}_{\rm abs}^{(0) {\rm T}} + \delta_{3}^{\prime\prime \star}\left(\omega\right) \, \mathcal{B}_{\rm abs}^{(0)} \mathcal{E}_{\rm abs}^{(0) {\rm T}} \right] {\bf e}^{(0)} = \delta n \left(\omega\right) \, {\bf e}^{(0)} \,,
\end{equation}
where 
\begin{eqnarray}
\delta_{1}^{\prime\prime}\left(\omega\right) + \delta_{2}^{\prime\prime}\left(\omega\right) & = & \delta_{1} + \delta_{2} + \Delta_{2,{\rm ax}}\left(\omega\right) \,, \nonumber \\
\delta_{1}^{\prime\prime}\left(\omega\right) - \delta_{2}^{\prime\prime}\left(\omega\right) & = & \left(\delta_{1} - \delta_{2} - \Delta_{2,{\rm ax}}\left(\omega\right)\right) {\rm cos}\left(2\chi\right) \,, \nonumber \\
\delta_{3}^{\prime\prime}\left(\omega\right) & = & \delta_{3} \, {\rm cos}\left(2\chi\right) + i \, \Delta_{3, {\rm ax}}\left(\omega\right) \, {\rm sin}\left(2\chi\right) \,.
\label{eq:delta_prime_prime}
\end{eqnarray}
As a matrix equation, this is simply
\begin{equation}
\Bigg[ \begin{array}{cc} \delta_{1}^{\prime\prime}\left(\omega\right) & \delta_{3}^{\prime\prime}\left(\omega\right) \\ \delta_{3}^{\prime\prime \star}\left(\omega\right) & \delta_{2}^{\prime\prime}\left(\omega\right) \end{array} \Bigg] \left[ \begin{array}{c} e^{(0)}_{\mathcal{E}} \\ e^{(0)}_{\mathcal{B}} \end{array} \right] = \frac{\delta n \left(\omega\right)}{\frac{1}{4} \left|\mathcal{E}_{\rm abs}^{(0)}\right|^{2}} \left[ \begin{array}{c} e^{(0)}_{\mathcal{E}} \\ e^{(0)}_{\mathcal{B}} \end{array} \right] \,.
\label{eq:eigenvalue_matrix_eqn_axions}
\end{equation}
The eigenvalues of the matrix on the left-hand side of this equation are calculated as before, and the refractive index changes $\delta n_{\pm}$ are again given by Eq.~(\ref{eq:delta-n_optical}) with the now frequency-dependent factors $\delta_{\pm}^{\prime\prime}\left(\omega\right)$ taking the form
\begin{eqnarray}
\!\!\!\!\!\! \delta_{\pm}^{\prime\prime} \left(\omega\right) & = & \frac{1}{2}\left( \delta_{1}^{\prime\prime}\left(\omega\right) + \delta_{2}^{\prime\prime}\left(\omega\right) \pm \sqrt{\left(\delta_{1}^{\prime\prime}\left(\omega\right) - \delta_{2}^{\prime\prime}\left(\omega\right)\right)^{2} + \left| 2 \delta_{3}^{\prime\prime}\left(\omega\right)\right|^{2}} \right) \nonumber \\
& = & \frac{1}{2} \left( \delta_{1} + \delta_{2} + \Delta_{2,{\rm ax}}\left(\omega\right) \pm \sqrt{ \left[\left(\delta_{1}-\delta_{2}-\Delta_{2,{\rm ax}}\left(\omega\right)\right)^{2} + \left(2 \delta_{3}\right)^{2}\right] {\rm cos}^{2}\left(2\chi\right) + \left(2\Delta_{3,{\rm ax}}\left(\omega\right)\right)^{2} \, {\rm sin}^{2}\left(2\chi\right) } \right) \,.
\label{eq:delta_axions}
\end{eqnarray}
In the EH model with a linearly polarized pump, the only change with respect to previous sections is that $\delta_{2} \to \delta_{2} + \Delta_{2,{\rm ax}}\left(\omega\right)$.  This is in agreement with Eq.~(13) of~\cite{Baldenegro-et-al-2019} (an apparent difference by a factor of 4 arising only because of different definitions of the pump amplitude).

We briefly mention here that the group index $n_{g} = n + \omega \, {\rm d}n/{\rm d}\omega$, though generally complicated, can be fairly easily calculated in the limits of linear and circular polarization.  Although $n < 1$ for some frequencies, we find (at least in these two limits) that $n_{g} > 1$ for {\it all} frequencies, so that relativistic causality is respected (as explained in footnote~\ref{fn:SR}).

\subsection{Eigenpolarizations}

Generally speaking, the matrix on the left-hand side of Eq.~(\ref{eq:eigenvalue_matrix_eqn_axions}) is hermitian but {\it not} symmetric, thanks to the imaginary contribution to the off-diagonal component $\delta_{3}^{\prime\prime}\left(\omega\right)$.  Unlike in previous sections, this means that its eigenvectors are generally complex, which in turn means that the eigenpolarizations are elliptically polarized.  Two angles are required to describe these eigenpolarizations: an ellipticity angle $\psi \in \left[ -\pi/4 \,,\, \pi/4 \right]$ giving the degree of elliptical polarization, and a rotation angle $\varphi \in \left( -\pi/2 \,,\, \pi/2 \right]$ giving the orientation of the major axes of the ellipses traced by ${\bf e}(t)$ and ${\bf b}(t)$ with respect to the basis $\left\{ \hat{\mathcal{E}} \,,\, \hat{\mathcal{B}} \right\}$.  The (normalized) eigenvectors thus form the columns of a unitary matrix $U\left(\psi \,,\, \varphi\right)$ which can be decomposed as follows:
\begin{equation}
U\left(\psi\,,\,\varphi\right) = E\left(\psi\right)  R\left(\varphi\right) = \left[ \begin{array}{cc} {\rm cos}\psi & i \, {\rm sin}\psi \\ i \, {\rm sin}\psi & {\rm cos}\psi \end{array} \right] 
\left[ \begin{array}{cc} {\rm cos}\varphi & -{\rm sin}\varphi \\ {\rm sin}\varphi & {\rm cos}\varphi \end{array} \right] \,.
\label{eq:U_varphi_and_psi}
\end{equation}
The matrix on the left-hand side of Eq.~(\ref{eq:eigenvalue_matrix_eqn_axions}) can be written as $U\left(\psi\,,\,\varphi\right) D U^{-1}\left(\psi\,,\,\varphi\right)$, where $D$ is a diagonal matrix whose entries are the eigenvalues $\delta_{+}^{\prime\prime}$ and $\delta_{-}^{\prime\prime}$ of Eq.~(\ref{eq:delta_axions}).  As before, we take $\delta_{+}^{\prime\prime}$ to be the first diagonal component of $D$, so that the left column of $U$ corresponds to the polarization with the larger refractive index change $\delta n_{+}$.  Explicit calculation of this form of the matrix yields the following relations, which determine $\psi$ and $\varphi$:
\begin{eqnarray}
\left(\delta_{1}-\delta_{2}-\Delta_{2,{\rm ax}}\left(\omega\right)\right) \, {\rm cos}\left(2\chi\right) & = & \left(\delta_{+}^{\prime\prime}\left(\omega\right)-\delta_{-}^{\prime\prime}\left(\omega\right)\right) {\rm cos}\left(2\varphi\left(\omega\right)\right) {\rm cos}\left(2\psi\left(\omega\right)\right) \,, \nonumber \\
-2 \Delta_{3,{\rm ax}}\left(\omega\right) \, {\rm sin}\left(2\chi\right) & = & \left(\delta_{+}^{\prime\prime}\left(\omega\right)-\delta_{-}^{\prime\prime}\left(\omega\right)\right) {\rm cos}\left(2\varphi\left(\omega\right)\right) {\rm sin}\left(2\psi\left(\omega\right)\right) \,, \nonumber \\
2 \delta_{3} \, {\rm cos}\left(2\chi\right) & = & \left(\delta_{+}^{\prime\prime}\left(\omega\right)-\delta_{-}^{\prime\prime}\left(\omega\right)\right) {\rm sin}\left(2\varphi\left(\omega\right)\right) \, .
\label{eq:varphi_and_psi}
\end{eqnarray}
Whilst in general the solutions $\psi\left(\omega\right)$ and $\varphi\left(\omega\right)$ 
are quite complicated, they become rather simple in the two polarization limits of the pump.  When the pump is linearly polarized ($\chi = 0$), we find $\psi = 0$ so that the probe polarizations are also linearly polarized; the only non-triviality is in their rotation with respect to the pump fields, the rotation angle being given by Eqs.~(\ref{eq:pol_rotation}) with $\delta_{2} \to \delta_{2} + \Delta_{2,{\rm ax}}\left(\omega\right)$.  On the other hand, for a circularly polarized pump ($\chi = \pm \pi/4$), we find $\psi = \pm \pi/4$ too, i.e., the eigenpolarizations of the probe are themselves circularly polarized. 
In the high-frequency regime where $\Delta_{3,{\rm ax}}\left(\omega\right) < 0$, it is straightforward to show~\footnote{There is a slight complication as there exists a degeneracy in $\varphi$ and $\psi$ (as defined by Eqs.~(\ref{eq:varphi_and_psi})) when $\chi = \pm \pi/4$: either $\varphi = 0$ and $\psi = \chi$, or $\varphi = \pi/2$ and $\psi = -\chi$.  However, plugging these solutions into Eq.~(\ref{eq:U_varphi_and_psi}), we find that switching between them simply amounts to multiplication of the columns of $U(\psi\,,\,\varphi)$ by $\pm i$, so that the eigenvectors remain unchanged (up to an unimportant phase).} that the larger refractive index change $\delta n_{+}$ is felt by the probe rotating in the same sense as $\mathcal{E}(t)$ and $\mathcal{B}(t)$, or equivalently (since in Fig.~\ref{fig:waves} the probe propagates {\it into} the page) by the probe with opposite handedness to that of the pump.

\subsection{Discussion of key axionic effects}

The terms $\Delta_{2,{\rm ax}}(\omega)$ and $\Delta_{3,{\rm ax}}(\omega)$ describe two key effects induced by the coupling to axions.  Moreover, each is dominant in a particular regime, for we have 
\begin{equation}
\frac{\Delta_{3,{\rm ax}}\left(\omega\right)}{\Delta_{2,{\rm ax}}\left(\omega\right)} = 2 \, {\rm cos}^{2}\frac{\theta}{2} \, \frac{\omega_{0}\omega}{\omega_{\rm ax}^{2}} \,.
\label{eq:axion_regimes}
\end{equation}
Therefore, whether $\omega_{0}\omega/\omega_{\rm ax}^{2}$ is large or small compared to $1$ determines which of the two terms is dominant.  (Both can be considered large when $\omega_{0}\omega/\omega_{\rm ax}^{2} \approx 1$, but then we are close to resonance, and the analysis performed here will cease to be valid.)

\subsubsection{Renormalization of $\delta_{2}$}

First, the parameter $\delta_{2}$ entering the local part of the effective Lagrangian is renormalized in a frequency-dependent way to become $\delta_{2} + \Delta_{2,{\rm ax}}\left(\omega\right)$.
When $\omega_{0}\omega/\omega_{\rm ax}^{2}$ is very large, it is clear from Eqs.~(\ref{eq:axion_contributions}) that $\Delta_{2,{\rm ax}}\left(\omega\right) \to 0$.  We may thus focus on the opposite limit, 
$\omega_{0}\omega/\omega_{\rm ax}^{2} \to 0$, 
where 
the Compton wavelength of the axion is much smaller than the typical photon wavelengths (or indeed when the background fields are static).
In this case, we have %
$\Delta_{2,{\rm ax}}(\omega) \to \eta^{2}/k_{\rm ax}^{2}$, with no residual dependence on frequency.
This is the limit in which the effective photon/photon interaction becomes local; $\eta^{2}/k_{\rm ax}^{2}$ can be absorbed into the definition of $\delta_{2}$, and the analysis of Sec.~\ref{sec:optical_Kerr} carries through as before.
We are thus provided with an explicit demonstration of the results of the local effective theory emerging in the correct limit.

For the first experimental probes of NLED, this renormalization is only expected to be significant if it is comparable to the ``bare'' EH value, $\delta_{2}^{\rm (EH)}$ (see Eqs.~(\ref{eq:EH-coeffs})).
In units with $\hbar = c = 1$, this requires $\eta/m_{\rm ax} \gtrsim \sqrt{\delta_{2}^{\rm (EH)}} \approx 3 \times 10^{4} \, {\rm GeV}^{-2}$.
This has not yet been ruled out by current experimental tests of axions but, since values of $\eta$ larger than about $10^{-10}\,{\rm GeV}^{-1}$ have been essentially excluded (see Fig.~111.1 of~\cite{PDGReview2018}), this would require an axion mass smaller than a few times $10^{-6} \, {\rm eV}$.  This makes it less likely that the condition $\omega_{0}\omega/\omega_{\rm ax}^{2} \gg 1$ will be satisfied, unless the background field is essentially static.  For the DeLLight experiment~\cite{Sarazin-et-al-2016}, which uses optical wavelengths with $\hbar\omega \sim 1\,{\rm eV}$ for both the background and the probe fields, this renormalization of $\delta_{2}$ is unlikely to be significant.

\subsubsection{Elliptical birefringence}

The second key axionic effect is the 
contribution $\propto \Delta_{3,{\rm ax}}\left(\omega\right)$ to the off-diagonal terms in Eq.~(\ref{eq:eigenvalue_matrix_eqn_axions}). 
Interestingly, this contribution is imaginary, and leads to probe eigenstates which are 
elliptically polarized. 
It is clear that $\Delta_{3,{\rm ax}}\left(\omega\right) \to 0$ when $\omega_{0}\omega/\omega_{\rm ax}^{2} \to 0$, as it should, since we then recover the results of the local effective theory where (as we have seen in Sec.~\ref{sec:optical_Kerr}) the relevant matrix is purely real.
In the opposite limit, 
we have already seen (in Eq.~(\ref{eq:axion_regimes})) that
$\Delta_{3,{\rm ax}}\left(\omega\right)$ becomes the dominant 
signature of the axion coupling, 
and yet, since it 
enters Eqs.~(\ref{eq:delta_prime_prime}) with a factor of ${\rm sin}\left(2\chi\right)$, it completely drops out when the pump is linearly polarized. 
We thus conclude that, 
when working in the limit $\omega_{0}\omega/\omega_{\rm ax}^{2} \gg 1$,
a circularly polarized pump is much more efficient than a linearly polarized one at inducing an axionic signature in the refractive index.
This result is made all the more interesting by the fact that this axionic signature includes birefringence, since (as seen in Sec.~\ref{sub:NLindex}) models with local effective photon/photon interactions show no birefringence at all when the pump is circularly polarized.

Working in the appropriate limit where $\omega_{0}\omega/\omega_{\rm ax}^{2}$ is very large, 
\begin{equation}
\Delta_{3, {\rm ax}}(\omega) \to -\frac{\eta^{2} c^{2}}{8} \frac{1}{{\rm cos}^{2}\frac{\theta}{2} \, \omega_{0}\omega} \,,
\end{equation}
and we see explicitly that this limiting case is independent of the actual value of $m_{\rm ax}$.
Moreover, in the optimized scenario with $\theta = 0$ and $\chi = \pm \pi/2$ (i.e., a counter-propagating probe and a circularly polarized pump), we have from Eq.~(\ref{eq:delta_axions}) that the birefringence $\delta_{+}-\delta_{-} = 2 \, \left| \Delta_{3,{\rm ax}}(\omega) \right|$.
It is appropriate to compare this with $\delta_{2}^{\rm (EH)}-\delta_{1}^{\rm (EH)}$, the birefringence predicted by the EH model (in the absence of axions) for a linearly polarized pump.
In units with $\hbar = c = 1$, they are comparable when $\eta/\sqrt{\omega_{0}\omega} \sim 2 \sqrt{\delta_{2}^{\rm (EH)}-\delta_{1}^{\rm (EH)}} \approx 4 \times 10^{4} \, {\rm GeV}^{-2}$.
At optical frequencies (as used in the DeLLight experiment~\cite{Sarazin-et-al-2016}), and given that current exclusion plots indicate $\eta$ is at most $\sim 10^{-10}\,{\rm GeV}^{-1}$ (see Fig.~111.1 of~\cite{PDGReview2018}), this circular birefringence turns out to be at least six orders of magnitude smaller than the linear birefringence of the EH model.

\section{Summary and conclusion
\label{sec:conclusion}}

Starting from the most general Lagrangian for electromagnetic fields consistent with Lorentz invariance, locality of effective interactions, and weak nonlinearities (so that the lowest nonlinear contributions are sufficient to describe the physics), we have derived an effective medium description for the propagation of weak probe waves in the presence of strong background fields, whose effects on the probe can be incorporated through well-defined electric, magnetic and magnetoelectric susceptibility tensors.  This description allows the assignment of a refractive index to the effective medium, though the index typically exhibits anisotropy and birefringence.  In the case where the background fields are provided by an intense propagating wave or ``pump'', there is a further dependence of the refractive index on the degree of elliptical polarization of the pump.  The effects of wave vector direction and polarization turn out to be neatly separated: the misalignment of the wave vectors of pump and probe is equivalent to an overall reduction in the pump intensity by a factor of $\mathrm{cos}^{4}\left(\theta/2\right)$, where $\theta$ is the tilt angle between the two wave vectors; and the ellipticity angle $\chi$ enters into the strength of the birefringence with a factor of $\mathrm{cos}\left(2\chi\right)$, being maximum for a linearly polarized pump and vanishing for a circularly polarized one.  Finally, factoring out the intensity of the pump allows us to extract the nonlinear 
index of vacuum, which in the Euler-Heisenberg model derived from QED is typically on the order of $10^{-33}\,{\rm cm}^{2}/{\rm W}$, around $18$ orders of magnitude smaller than in nonlinear optical media.

Locality of effective photon/photon interactions was relaxed as a constraint through coupling the electromagnetic field to an axion field of unspecified mass.  Generation of an axion field by the interaction between pump and probe, followed by backreaction of the axion field on the probe, yields a contribution to the effective photon/photon coupling which is dispersive when the axion's Compton wavelength is larger than or of the same order as the typical photon wavelength.  In the case of a linearly polarized pump, this amounts to a straightforward renormalization of one of the post-Maxwellian parameters entering the Lagrangian, but when the pump is circularly polarized, there exists a residual birefringence that (as mentioned above) would vanish if the effective photon/photon interactions were purely local.  Whether the axionic contribution to the birefringence is larger for a linearly or circularly polarized pump depends on the typical photon wavelength (defined via the geometric mean of the pump and probe frequencies): if it is much smaller than the Compton wavelength of the axion, the effect is larger for a circularly polarized pump; conversely, if the typical photon wavelength is much larger than the Compton wavelength of the axion, the effect is larger for a linearly polarized pump.

Concerning the DeLLight experiment~\cite{Sarazin-et-al-2016}, whose aim is to detect the deflection of a probe wave by the index variation induced by a tightly focused laser pulse, the results presented here confirm that an effect of this kind should indeed be seen,
and any effects due to axions are expected to be subdominant.  %
However, as discussed after Eqs.~(\ref{eq:L-vectors}) (particularly in footnote~\ref{fn:DeLLight}), we have shown that only a single pump pulse is required, rather than two counter-propagating pump pulses as proposed in~\cite{Sarazin-et-al-2016}.  
We conclude that the proposal presented in~\cite{Sarazin-et-al-2016} can be simplified by keeping only the pump pulse which is counter-propagating with respect to the probe.  We are currently performing numerical simulations of the 
simplified %
DeLLight experiment, whose results will be published in a future work.
It is worth noting that the two-pump proposal of~\cite{Sarazin-et-al-2016} would be an interesting case in which four-wave mixing processes could turn out to be important, as the two counter-propagating pump pulses would engender a stationary oscillation in space that could act as a 
diffraction grating for the probe~\cite{DiPiazza-et-al-2006}; this, however, is beyond the scope of the present work. %

\section*{Acknowledgments}

I thank Xavier Sarazin and Fran\c{c}ois Couchot for many fruitful discussions on the DeLLight experiment, which motivated the present work, and for their careful reading of and constructive comments on the manuscript.
I thank Renaud Parentani for many discussions on the theoretical side, and I am grateful to Laurent Schoeffel for a very useful exchange on how to include the coupling to axions. 
This work was mainly done during two short-stay postdocs at LAL in 2018, funded by LAL and LPT (Laboratoire de Physique Th\'{e}orique).  It is continuing to be supported by the French National Research Agency via the grant no. ANR-18-CE31-0005-01.

\bibliography{biblio}

\end{document}